\documentclass[preprint,sort&compress]{elsarticle}
\usepackage[utf8]{inputenc}
\usepackage[english]{babel}
\usepackage{wrapfig}
\usepackage{amsmath}
\usepackage{amsfonts}
\usepackage{amssymb}
\usepackage{makeidx}
\usepackage{graphicx}
\usepackage{lipsum}
\usepackage{lmodern}
\usepackage{setspace}
\usepackage{fourier}
\usepackage[left=2cm,right=2cm,top=2cm,bottom=2cm]{geometry}\usepackage[utf8]{inputenc}
\usepackage{subcaption}
\usepackage{comment}
\usepackage{color}

\definecolor{byzantium}{rgb}{0.44, 0.16, 0.39}

\begin{document}

\title{Material vs. structure: Topological origins of band-gap truncation resonances in periodic structures}


\author{Matheus I. N. Rosa \fnref{fn2}}
\author{Bruce L. Davis \fnref{fn1}}
\author{Liao Liu \fnref{fn1}}
\author{Massimo Ruzzene \fnref{fn2,fn1}}
\ead{massimo.ruzzene@colorado.edu}
\author{Mahmoud I. Hussein \fnref{fn1,fn3}}
\ead{mih@colorado.edu}

\fntext[fn2]{Paul M. Rady Department of Mechanical Engineering}
\fntext[fn1]{Ann and H.J. Smead Department of Aerospace Engineering Sciences}
\fntext[fn3]{Department of Physics}

\address{University of Colorado Boulder, Boulder, CO, 80303}

\begin{abstract}
While resonant modes do not exist within band gaps in infinite periodic materials, they may appear as in-gap localized edge modes once the material is truncated to form a finite periodic structure. Here, we provide an analysis framework that reveals the topological origins of truncation resonances, elucidating formally the conditions that influence their existence and properties. Elastic beams with sinusoidal and step-wise property modulations are considered as classical examples of periodic structures. Their non-trivial topological characteristics stem from the consideration of a phason parameter that produces spatial shifts of the property modulation while continuously varying how the boundaries are truncated. In this context, non-trivial band gaps are characterized by an integer topological invariant, the Chern number, which is equal to the number of truncation resonances that traverse a band gap as the phason is varied. We highlight the existence of multiple chiral edge states that may be localized at opposite boundaries, and illustrate how these can be independently tuned by modified boundary-specific phason parameters. Boundary phasons modify the truncation of only one boundary at a time. Furthermore, we show that the frequency location of a truncation resonance is influenced by the modulation wavelength, modulation volume fraction, boundary conditions, and number of cells comprising the finite structure, thus quantifying its robustness to these factors. Non-topological in-gap resonances induced by a defect are also demonstrated, with their frequency dependence on the phason investigated to elucidate their contrast to truncation resonances. A coupling between topological and non-topological modes is shown to be possible when the defect is located at an edge. Finally, experimental investigations on bi-material phononic-crystal beams are conducted to support these findings. Our results provide a fundamental perspective on the topological character of truncation resonances in periodic structures and how this character relates to the underlying periodic material properties. The tunability of these unique structural resonances through material-property modulation may be exploited both in applications where in-gap resonances are not desired, such as vibration attenuation and thermal conductivity reduction, or where in-gap resonances provide a functional role, such as filtering, waveguiding, energy harvesting, and flow control by phononic subsurfaces.

\end{abstract}

\begin{keyword}
Phononic materials \sep band-gap resonances  \sep topological protection \sep phasons \sep experimental phononics
\end{keyword}

\date

\maketitle

\section{Introduction}
\label{Introd}

The study of elastic wave propagation in a continuous periodic medium is a classical problem in mechanics that can be traced back to Rayleigh in 1887~\cite{Rayleigh1887}. With the advent of composite materials, the interest in this problem surged with early contributions in the 1950s~\cite{Rytov_1956} and 1960s~\cite{Sun_1968} formulating dispersion relations for wave propagation in laminated composites, and other forms of periodic media~\cite{Heckl_1964,Mead_1970}, followed by extension to multi-dimensional composites in the 1970s~\cite{Nemat-Nasser_1975}. The field re-emerged in the early 1990s with the study of phononic crystals~\cite{Sigalas_1993,Kushwaha_1993} and the establishment of formal connections with lattice dynamics in crystals~\cite{Maradudin_1969}, and gathered further pace with the rise of acoustic and elastic metamaterials~\cite{liu2000locally}. In all these studies, periodicity is utilized enabling dynamic characterization by considering a representative unit cell, as commonly done in condensed matter physics~\cite{Kittel_1976}. Calculating the dispersion relation, or the band structure, using the Floquet/Bloch theorem~\cite{Floquet_1883,Bloch1929} formally enforces the assumption of an extended medium with an~\textit{infinite} number of unit cells. This is not only computationally rewarding, but physically provides a fundamental description of the modal wave propagation properties of the medium under investigation$-$removing any influence of overall size and external boundary conditions.~In this framework, the medium under consideration is rendered a \textit{material} with characteristic \textit{intrinsic} properties, such as band gaps (whose locations may be predicted analytically~\cite{Mead_1975a,Mead_1975b,Xiao_2011,Raghavan2013}) and other key features revealed by the nature of the band structure. The thermal conductivity, for example, is an intrinsic material property that is directly influenced by the band structure$-$determined by analysis of only a single atomic-scale unit cell~\cite{Callaway_1959,Holland_1963}. Effective dynamic properties, such as effective density and Young's modulus~\cite{Nemat-Nasser_2011}, provide another example of intrinsic material properties.~On the other hand, unless a medium practically comprises thousands or millions of unit cells (as in a bulk crystal for example), realistic realizations are formed from a relatively small \textit{finite} number of unit cells, yielding a periodic \textit{structure}, rather than a material, with \textit{extrinsic} properties. This is particularly the case in engineering problems such as sound~\cite{SanchezPerez1998} and vibration~\cite{Vasseur2001} isolation, and other similar applications~\cite{hussein2014,Jin_2021}, and also the case in nanoscale thermal transport~\cite{Maruyama_2002} where unique dynamical properties emerge primarily from the presence of finite size along the direction of transport.

\subsection{Truncation resonances}
A periodic structure in practice may still consist, in some cases, of a relatively large but tractable number of unit cells, and in other cases, of only a few unit cells along the direction of vibration transmission.~The number of cells impacts the degree of attenuation within a band gap~\cite{Hussein2006Analysis}.~However, the contrast between the material and structure behavior may not be limited to only quantitative differences but also to fundamental qualitative distinctions. 
One noticeable anomaly between the material and structure responses is the possibility of existence of resonances inside band gaps, i.e., resonance peaks in the frequency response function (FRF) of a finite periodic structure that appear within band-gap frequency ranges of the corresponding infinite periodic material. These resonances are often referred to as \textit{truncation resonances}~\cite{hussein2015flow,Al-Babaa_2019} because they emerge from the truncation of a medium that is otherwise formed from an infinite number of unit cells.~These resonances are associated with mode shapes that localize at the truncation junction, and are thus also commonly referred to as \textit{edge} or \textit{surface modes}~\cite{wallis1957effect,Camley1983transverse,Djafari_1983,Boud1993surface,Boudouti_1995theory,Boudouti1996theory,Hladky2005localized,Ren2007theory,Shuvalov2018}.~The presence of these modes has been uncovered theoretically by Wallis~\cite{wallis1957effect} in his study of a finite discrete diatomic chain of atoms with free ends.~This followed the work of Born on finite atomic chains~\cite{Born_1942} which was motivated by the study of the influence of lattice vibrations on X-ray scattering.~Recent studies extended Wallis' theory of finite discrete chains to more general conditions~\cite{Al-Babaa_2017_Poles,Al-Babaa_2019,Bastawrous_2022} and experiments on chains of discrete-like coupled spheres validated the theory~\cite{Hladky2005localized}.  

The problem of truncation resonances in continuous periodic media$-$the focus of this paper$-$has also been investigated extensively.~Early studies examined one-dimensional wave propagation in periodically layered/lamenated composites, also referred to as superlattices. Existence conditions for truncation resonances were derived for semi-infinite superlattices for out-of-plane~\cite{Camley1983transverse,Boud1993surface,Boudouti1996theory} and in-plane~\cite{Djafari_1983,Boudouti_1995theory,Shuvalov2018} waves.~It was shown that surfaces modes in some instances may appear below the lowest bulk band, i.e., the band that hosts conventional resonances.~Investigations of the truncation phenomenon were also done on finite layered phononic crystals examining transverse waves~\cite{Ren2007theory,ren2010surface}, on finite beam-based phononic crystals~\cite{DavisIMECE2011,hvatov2015free} and locally resonant elastic metamaterials~\cite{xiao2013flexural,Sangiuliano_2020}, and on rod-based phononic crystals~\cite{Cameiro_2021,AlBabaa_2022}.~Among the factors that influence the frequency location of the truncation resonances are the unit-cell symmetry and the boundary conditions~\cite{ren2010surface,hvatov2015free,Sangiuliano_2020,Cameiro_2021,AlBabaa_2022}.~When there is more than one layer in the unit cell, the number of surface states increases~\cite{Boudouti1996theory,Shuvalov2018}.~Techniques proposed for control of the truncation resonances also include tuning of unit-cell spatial material distribution or volume fraction~\cite{DavisIMECE2011}, and the anomalous addition of a ``cap layer"~\cite{Boud1993surface,Boudouti1996theory} or a ``tuning layer"~\cite{Hussein_JSV_2007,DavisIMECE2011} at the edge of the structure.~A cap layer is simply a homogeneous layer, whereas a tuning layer is a purposefully truncated single unit cell.~The concept of truncation resonances is also relevant to other areas in applied physics such as photonic crystals~\cite{yeh1978optical} and quantum lattices~\cite{ren2010surface}.

\subsection{Connection to topological physics}
The principle of a truncation resonance is fundamentally connected to the periodic structure's topological properties; this connection forms the core focus of the present study.~Inspired by the emergence of topological insulators in condensed matter physics~\cite{hasan2010colloquium}, classical analogues have been developed in photonics~\cite{ozawa2019topological} and phononics~\cite{miniaci2021design}, demonstrating the features of robust topological waves. In passive elastic materials, topological interface modes are created by contrasting two materials with band gaps existing at the same frequencies, but characterized by different topological invariants. Examples include interface modes in one-dimensional (1D) structures~\cite{pal2017edge,chaunsali2017demonstrating,wang2020robust,yin2018band} in analogy to the Su-Schrieffer-Heeger model~\cite{su1979solitons}, and waveguiding along interfaces in two-dimensional (2D) materials in analogy to the Quantum Spin Hall Effect~\cite{susstrunk2015observation,miniaci2018experimental} or to the Quantum Valley Hall Effect~\cite{torrent2013elastic,pal2017edge,vila2017observation}. These effects rely on symmetry breaking by interfacing two domains whose unit cells have opposite symmetries, which results in contrasting topological properties in the reciprocal space. Hence, an actual interface between two materials is required, which presents a contrast to the truncation resonances we explore in this paper. We will show an intriguing connection that stems from a stronger type of topological effect associated with the Quantum Hall Effect (QHE)~\cite{thouless1982quantized,klitzing1980new}. The QHE manifests in 2D lattices of electrons under the presence of a strong magnetic field, which leads to robust edge waves that propagate along the boundaries of a finite sample (structure), without backscattering at corners or defects. It is therefore sufficient to exploit the interface between a single material medium and vacuum. However, such a strong effect requires breaking time reversal symmetry, which in the quantum case is achieved through the magnetic field. Emulating similar features on 2D elastic materials is possible through active components that break time reversal symmetry, such as rotating frames~\cite{wang2015coriolis} or gyroscope spinners~\cite{wang2015topological,nash2015topological}. An alternative that has emerged later, and which we adopt here, is to map the QHE to 1D passive structures that have extended dimensionality emanating from their parameter spaces~\cite{kraus2016quasiperiodicity,prodan2015virtual}. This has been achieved by using patterned mechanical spinners~\cite{apigo2018topological}, spring-mass lattices~\cite{rosa2019edge}, acoustic waveguides~\cite{apigo2019observation,ni2019observation}, and continuous phononic crystals or elastic metamaterials with modulations of inclusions such as ground springs~\cite{pal2019topological}, stiffeners~\cite{gupta2020dynamics}, and resonators~\cite{xia2020topological,rosa2021exploring}. In these examples, edge states localized at the boundaries of 1D periodic and quasi-periodic finite domains are observed to appear in correspondence to non-zero topological inavariants called \textit{Chern numbers}. The boundary at which the localization occurs can be determined by a phason parameter that is associated with spatial shifts in the medium’s modulated properties. This feature leads to possibilities for topological pumping by varying the phason parameter continuously along time~\cite{grinberg2020robust,chen2019mechanical,xia2021experimental} or along a second spatial dimension~\cite{rosa2019edge,riva2020edge}, inducing a transition of the edge states from being localized at one boundary to the other. Thus, energy can be "pumped" between two boundaries of a system through a transition of a topological edge state.~The application of the field of topology to elastic and acoustic material systems has been attracting much interest in recent years~\cite{ma2019topological,miniaci2021design,huang2021recent}.\\

In this paper, we provide a formal framework for the identification of the topological character of truncation resonances in periodic structures, drawing on concepts from the QHE. We consider a family of periodic elastic beams with either sinusoidal or step-wise property modulations. The modulations offer key parameters that expand the structure's property space and allow us to readily apply the concepts of topological band theory. In particular, the variation of a periodic beam's spectral properties with respect to the modulation wavelength allows us to extract the Chern numbers of the band-gaps and identify the locations of truncation resonances. Then, the phason parameters associated with spatial shifts of the modulations further characterize the truncation resonances as topological edge states spanning the band gaps. The frequency dependence of the location of a truncation resonance on the phason has recently been predicted, for periodic rods, by means of a closed-form transfer-matrix-based mathematical formulation~\cite{AlBabaa_2022}. Here, we investigate, for periodic flexural beams, the topological origins of this class of relations.  We show that the number of truncation resonances within a gap is equal to the predicted Chern number, for any set of boundary conditions, although the particular features of their branches as they traverse the gaps may vary. We elucidate how additional \textit{boundary phason} parameters can be defined, formalizing the notion of the tuning layer~\cite{Hussein_JSV_2007,DavisIMECE2011}, to manipulate the edge states localized at different boundaries independently. Furthermore, we examine the convergence of the truncation resonant frequencies as a function of the number of unit cells$-$a matter of significant practical importance especially when this number is relatively small. The fundamental differences, and the possibility of coupling, between truncation resonances and corresponding non-topological \textit{defect-mode} resonances are then investigated. Next, we provide laboratory results using a bi-material phononic-crystal beam as experimental validation of some of the key features of truncation resonances and their association with topological theory. Finally, we use our experiments to explore yet another important factor in the design space, namely the role of the materials' volume fraction within the unit cell in influencing the frequency locations of the truncation resonances.

The paper is organized as follows: following this introduction, Section \ref{Sec:ModBeams} provides a description of the considered periodic flexural beams and their boundary truncation through phasons. Next, Section \ref{Sec:Theory} develops the theory and computational analysis to characterize the topological properties of truncation resonances and those of non-topological defect resonances, and the coupling of the two types of resonances, followed by Section \ref{Sec:Experiments} which provides experimental results and further analysis. Finally, Section \ref{Sec:Disc} has a general discussion on the key findings and their broader implications to related areas of research, and Section~\ref{Sec:Conc} provides a closing summary and outlines possible future research directions.\\

\section{Modulated phononic-crystal beams: Truncation characterization by phasons}\label{Sec:ModBeams}

\begin{figure*}[b!]
\centering
\includegraphics{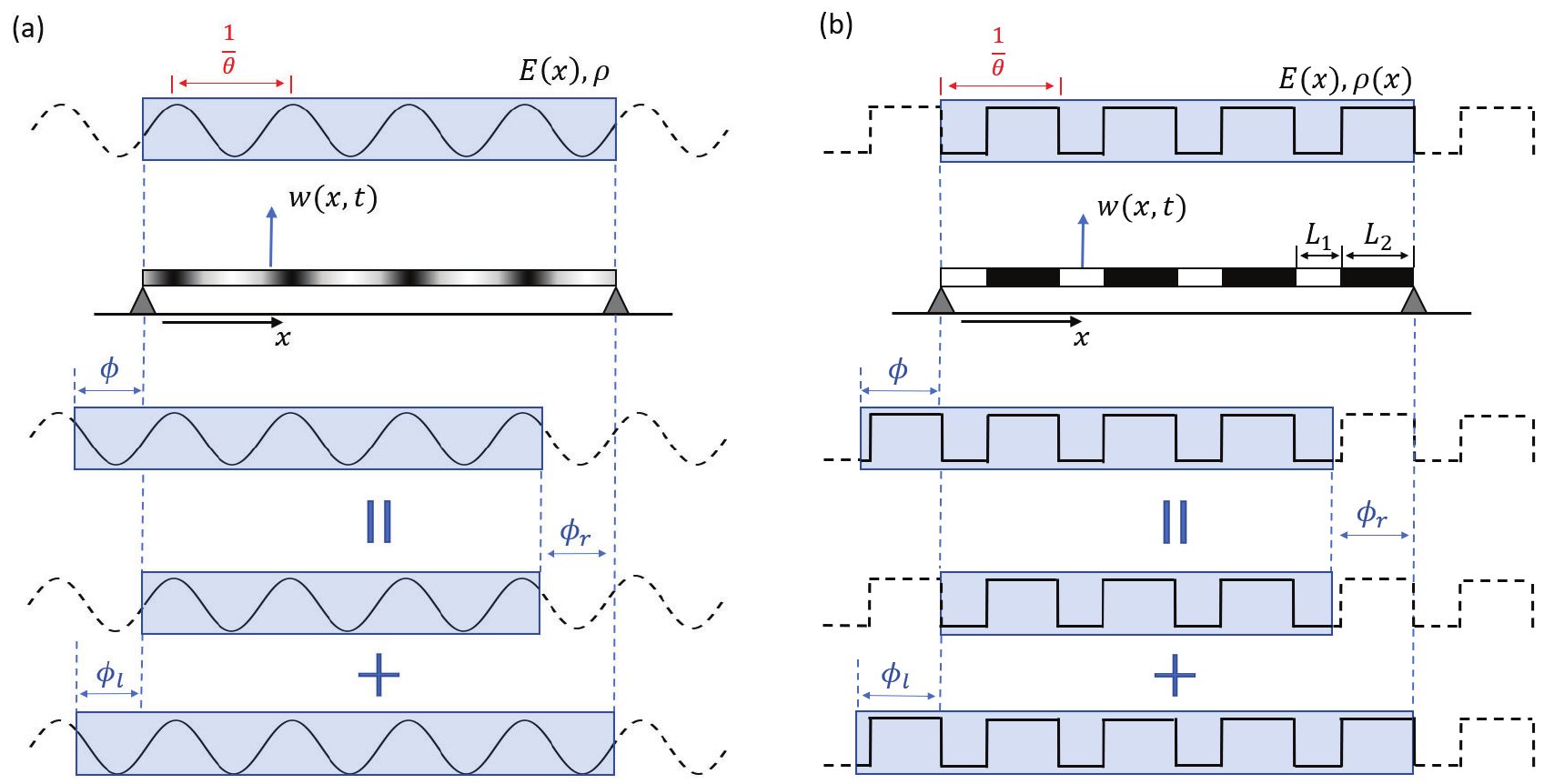}
\caption{\label{fig:fig1}{Elastic periodic beams with (a) sinusoidal and (b) step-wise property modulation whose spatial distribution is defined by a phason $\phi$ or boundary phasons $\phi_r$ and $\phi_l$. A modulation characterized by $\phi$ is a superposition of modulations characterized by $\phi_r$ and $\phi_l$.}
}         
\end{figure*}

We consider elastic beams undergoing flexural motion described by transverse displacement $w=w(x)$ and angle of rotation $\varphi=\varphi(x)$, where $x$ is the axial position, as classical examples of 1D periodic materials or structures. The properties of the beam are the Young's modulus $E=E(x)$, shear modulus $G=G(x)$, density $\rho=\rho(x)$, cross-sectional area $A=A(x)$, and second moment of area $I=I(x)$. These properties are modulated in space as illustrated in Fig.~\ref{fig:fig1}. Two scenarios are considered; in the first the Young's modulus is modulated according to a cosine function, i.e. $E(x)=E_0[1+\alpha\cos(2\pi\theta x - \phi)] $, while other parameters remain constant (Fig.~\ref{fig:fig1}(a)). This cosine-modulated phononic crystal (CM-PnC) serves as an idealized continuous periodic waveguide used to illustrate the behavior of interest in a simple setting. It is characterized by a unit cell of length $a=1/\theta$, where $\alpha$ is the amplitude of the modulation with respect to the mean value $E_0$ and $\theta$ may be viewed as the modulation wavenumber. The second case corresponds to a beam modulated in a step-wise fashion, which we refer to as step-wise modulated phononic crystal (SM-PnC). It generically represents a periodic material of two alternating layers of lengths $a_1$ and $a_2$, with different constituent material or geometrical (e.g. cross-sectional area) properties. In this case, the material or geometrical properties are modulated through a step-wise function of period $a=1/\theta=a_1+a_2$ that takes two different values in the intervals of length $a_1$ and $a_2$. 

The appearance of in-gap resonances stems from the truncation of the boundaries. The truncation details are here characterized by \textit{phason} parameters that are connected to non-trivial topological properties. The most natural choice of the phason is simply the phase $\phi$ of the property modulations, which rigidly shifts the modulation in space. Thus it results in a simultaneous change of the local properties of the beam at both boundaries. This is illustrated in the schematics of Fig.~\ref{fig:fig1} for both the sinusoidal and step-wise modulations. The blue boxes highlight the region of the modulations selected to form the properties of the finite beams. From a given initial configuration, a change in phason over the range $0<\phi<2\pi$ (higher values of $\phi$ do not need to be considered due to the periodicity) can be interpreted as simultaneously adding a segment of length $\phi a/2\pi$ to the left boundary, while removing the same length from the right boundary. This will naturally influence any vibration mode localized at either boundary. It's effect can be further understood as the superpostion of two independent parameters which we call boundary phasons. A change in the right boundary phason $\phi_r$ corresponds to removing a length $\phi_r a/2\pi$ from the right boundary while keeping the left boundary unchanged, while a change in the left boundary phason $\phi_l$ corresponds to adding a length $\phi_l a/2\pi$ to the left boundary while keeping the right boundary unchanged. Hence, changing the phason $\phi$ corresponds to changing both the left and right boundary phasons by the same amount (as illustrated in the figure). As we will show, the boundary phasons independently tune the topological truncation resonances at their respective boundary, and their superimposed effect leads to the variation of the resonances with respect to the conventional phason $\phi$.  

Herein, the flexural motion of the beam is modeled through Timoshenko theory as governed by the following two coupled equations:


\begin{subequations}
\label{TimoEq}
\begin{equation}\label{TimoEq1}
\rho A\frac{\partial^2 w}{\partial t^2}-q(x,t)=\frac{\partial}{\partial x} \left[ \kappa_{\mathrm{s}} AG \left( \frac{\partial w}{\partial x} -\varphi \right) \right] ,
\end{equation}
\begin{equation}\label{TimoEq2}
\rho I\frac{\partial^2 \varphi}{\partial t^2}=\frac{\partial}{\partial x} \left( EI \frac{\partial \varphi}{\partial x} \right) +\kappa_{\mathrm{s}} AG \left( \frac{\partial w}{\partial x} -\varphi \right) ,
\end{equation}
\end{subequations}

\noindent where $\kappa_{\mathrm{s}}$ denotes the shear coefficient, and $t$ and $q=q(x,t)$ represent time and the external forcing, respectively. Equations~\ref{TimoEq1} and~\ref{TimoEq2} are combined to yield a single fourth-order partial differential equation with only $w$ as the dependent variable~\cite{liu2012wave}. In our investigation, we consider three types of problems: a Bloch dispersion analysis problem for a unit-cell representing an infinite material, an eigenvalue analysis problem for a finite structure with arbitrary boundary conditions (BCs), and a harmonic forced-response problem for a finite structure with arbitrary BCs. In the first two problems, we set $q=0$ and 
\begin{equation}
\label{EigenEq}
w(x,t)=\hat{w}e^{i(\mu x-\omega t)},
\end{equation}
where $\omega$ denotes the frequency. In Eq.\ref{EigenEq}, we set $0 \leq \mu \leq \pi/a$ for the Bloch dispersion problem, where $\mu=0$ is used for the finite periodic-structure eigenvalue with arbitrary BCs. The results are obtained by a finite-element discretization of the equations of motion. The implementation details of these methods are omitted here for brevity since they are widely available in the literature (for example, see Ref.~\cite{phani2006wave}). 

Motivated by the experimental portion of this work (see Section \ref{Sec:Experiments}), we select the following parameters. The SM-PnC consists of a bi-material beam composed of alternating layers of Aluminum (Al) and the polymer acrylonitrile butadiene styrene (ABS). These materials are selected due to the contrast of mechanical properties leading to wide band gaps. Their properties are as follows: Young's moduli $E_{\mathrm{Al}}=68.9$ GPa and $E_{\mathrm{ABS}}=2.4$ GPa, shear moduli $G_{\mathrm{Al}}=25.9$ GPa and $G_{\mathrm{ABS}}=0.872$ GPa, and densities $\rho_{\mathrm{Al}}=2700$ kg/m$^3$ and $\rho_{\mathrm{ABS}}=1040$ kg/m$^3$, respectively. While we will allow the unit-cell length to vary through the $\theta$ parameter, the ABS polymer length filling fraction is fixed as $a_{\mathrm{ABS}}/a=0.2$; this ratio will be changed only in Section~\ref{CellVolF}. For purposes of comparison, the properties of the CM-PnC are then chosen to make it statically equivalent~\cite{Hussein2006Analysis} to the SM-PnC by selecting a fixed density $\rho_0=(0.2\rho_{\mathrm{ABS}}+0.8\rho_{\mathrm{Al}})$ and elastic modulus modulation with a mean value of $E_0=(0.2/E_{\mathrm{ABS}}+0.8/E_{\mathrm{Al}})^{-1}$. We consider a Poisson's ratio of $\nu=0.33$, which consequently determines the shear modulus through the relation $G=E/(2(1+\nu))$. Throughout this paper, the CM-PnC modulation amplitude is fixed at $\alpha=0.9$, and the beams have a square cross-section geometry with side length $h=2.54$cm. The finite-element analysis follows by discretizing the beams with linear Timoshenko beam elements with a shear coefficient of 5/6. The beam element length varies according to the case studied but does not exceed a maximum length of $\bar{a}/100$, where $\bar{a}=203$ mm is the unit-cell size of the experimental beams and is used as a reference unit-cell length throughout the paper.

\begin{figure*}[t!]
\centering
\includegraphics{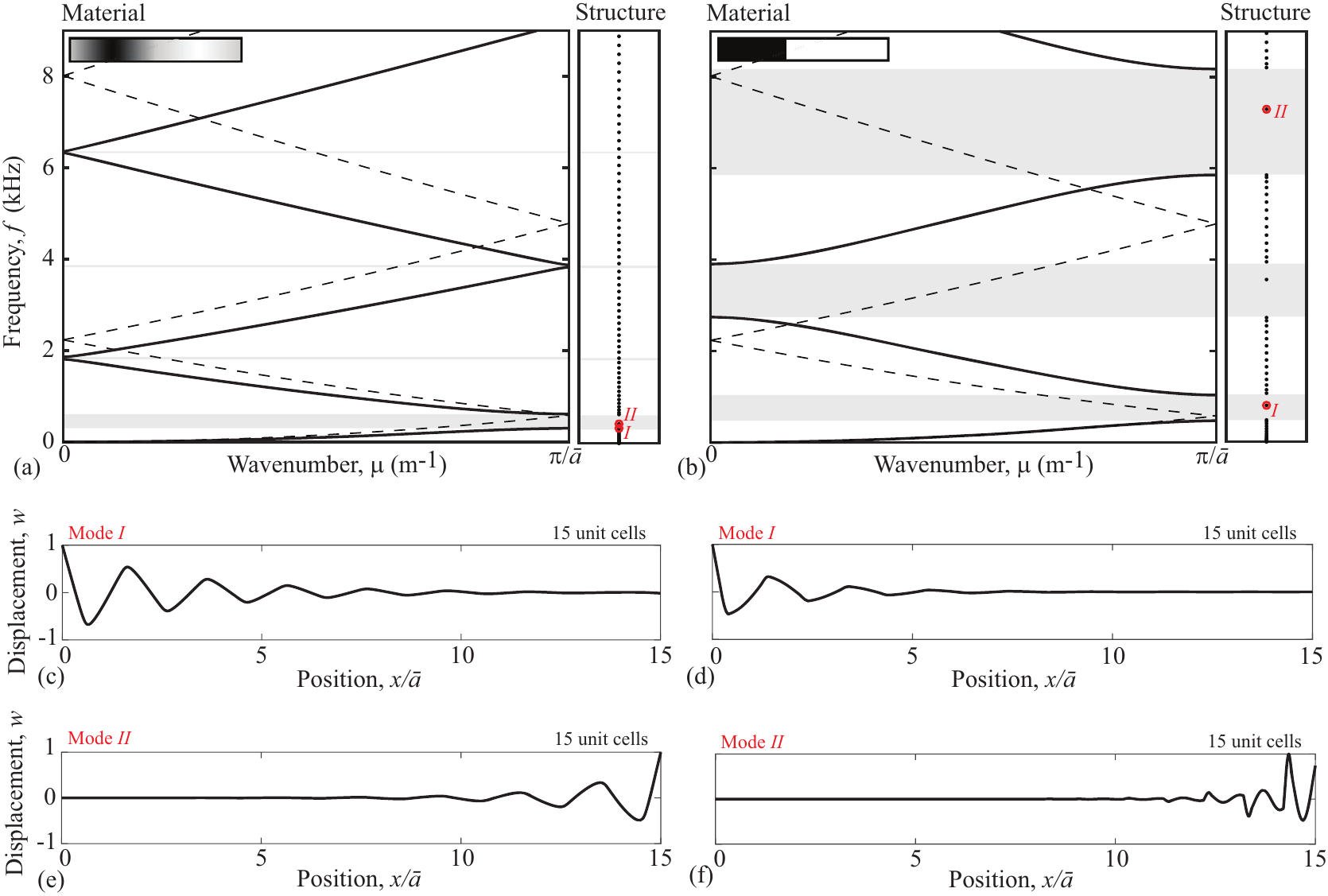}
\caption{\label{fig:fig2}Material versus structure properties. Dispersion diagrams (material) for the CM-PnC and the SM-PnC models are displayed in (a-b) as solid lines, while dashed lines correspond to the homogenized beam dispersion. Band-gap frequency ranges are shaded grey. A finite structure with 15 unit cells exhibits in-gap truncation resonances as illustrated alongside the dispersion diagrams, with selected mode shapes displayed in (c-f). For both models, the unit-cell length is $\bar{a}=203$ mm.}          
\end{figure*}

Figure~\ref{fig:fig2} presents a comparison between the properties of the CM-PnC and SM-PnC for the reference unit-cell size $\bar{a}=203$ mm, highlighting the contrast between material and structure. Panels (a) and (b) display their dispersion diagrams in a frequency range of interest from 0-9 kHz, which is a material feature. Both CM-PnC and SM-PnC exhibit the same long-wave static limit that approaches the dispersion of the homogenized beam with material property constants $\rho_0,E_0$ (dashed lines), but display different band-gaps (shaded gray regions). In particular, the SM-PnC has wider gaps due to its discrete nature and the contrast of both densities and elastic moduli, while the CM-PnC has smaller gaps due to a fixed density and a continuous variation of the elastic modulus only. On the right side of the dispersion diagrams, the eigenfrequencies of representative finite beams with 15 unit cells and free-free BCs are plotted as black dots, with $\phi=0.2\pi$ and $\phi=0.4\pi$ selected for the CM-PnC and SM-PnC beams, respectively. Truncation resonances are observed to appear in band gaps, a feature which is unique to the structure, non-existing at the material level. An arbitrary phason value is chosen here to produce a large number of truncation resonances as an example, but the behavior with the full range of $\phi$ will later be explored and explained. The truncation resonances are localized at one of the two boundaries of the finite beams, with selected mode shapes displayed in Figs.~\ref{fig:fig2}(c-f). By looking at such isolated cases (as has been largely done in previous studies), there is no apparent reason or pattern pertaining to the appearance of in-gap resonances, why are they localized at one boundary instead of the other, and why these features can change by selecting different BCs or different numbers of unit cells, etc. In the following sections, we will shed light on all of these questions by illustrating the topological character of in-gap truncation resonances associated with non-zero Chern numbers, and consequently how they can be manipulated through the phason and other parameters or design features. 

\section{Topological properties of modulated phononic-crystal beams}
\label{Sec:Theory}
In this section we develop the theoretical tools for the topological characterization truncation resonances by examining their behavior inside band gaps. We begin by investigating the effect of the modulation wavenumber $\theta$, which allows us to extract the topological invariants (Chern numbers). We then show how the Chern numbers are related to in-gap truncation resonances through the variation of the phason parameters. We also study the effect of the number of unit cells comprising the finite structure on the convergence of the truncation resonance frequencies. Finally, we provide a comparison between topological truncation resonances and non-topological defect resonances, highlighting their key differences and demonstrating the possibility of their coupling as a defect is moved towards a boundary. 

\subsection{Topological characterization by the Chern number}
In principle, the Chern number characterizes the topology of a vector field defined over a two-dimensional torus. For 2D periodic materials the torus is composed of two orthogonal wavenumber coordinates $\kappa_x$ and $\kappa_y$ and describes the reciprocal space Brillouin zone~\cite{hatsugai1997topological,susstrunk2015observation,miniaci2018experimental,pal2017edge,vila2017observation}. For 1D modulated materials such as the considered beams, the phason $\phi$ serves as an additional dimension and replaces the missing wavenumber component to form a torus based on $\kappa$ and $\phi$.~\cite{rosa2019edge}. The eigenvector field is the Bloch mode displacement $\hat{w}_n(\kappa,\phi)$ corresponding to the $n_{\text{th}}$ band defined over the torus $(\kappa,\phi)\in \mathbb{T}^2 = [0,2\pi]\times [0,2\pi]$, recalling that the dispersion is $2\pi$-periodic in both $\phi$ and $\kappa$, with $\kappa=\mu a$ defined as the non-dimensional wavenumber. Due to the continuous nature of the beams, the dispersion frequency bands are invariant with $\phi$, which only produces a shift in the choice of the unit cell. However, the variation of $\phi$ produces changes in Bloch eigenvectors, which may reflect in non-trivial topological properties. The Chern number $C_n$ for the $n_{\text{th}}$ band is defined as
\begin{equation}\label{Cherneq1}
C_n = \dfrac{1}{2 \pi i} \int_{\mathcal{D}} \beta_n \; d\mathcal{D},
\end{equation}
where $\mathcal{D}=\mathbb{T}^2$, $\beta_n=\nabla \times \mathbf{A}_n$ is called the Berry curvature, and $\mathbf{A}_n = \hat{w}_n^* \cdot \nabla \hat{w}_n$ is the Berry connection, with $()^*$ denoting a complex conjugate. The Chern number is an integer that quantifies the topological properties of the bands; these are robust to small perturbations in the system's unit cell as long as these perturbations do not close the gaps separating the bands. Among other features, the Chern number is related to discontinuities (or vorticities) in the eigenvector field~\cite{hatsugai1997topological}, localization of the Berry curvature~\cite{pal2017edge}, and to phase accumulation of the Bloch modes along cyclic paths in the torus Brillouin zone~\cite{thouless1982quantized,rosa2019edge}.

Of particular relevance to the present work is the bulk-boundary correspondence principle that relates the existence of in-gap edge states in finite systems to the Chern numbers~\cite{hatsugai1993chern}. This is done through the computation of a gap label $C_g$ given by the summation of the Chern numbers of the bands below the gap, i.e. $C_g^{(r)} = \sum_{n=1}^r C_n$, which is equal to the number of truncation resonances found inside such gap when the phase $\phi$ varies in an interval of $2\pi$ (see Section~\ref{ContTopStates} for more details). However, the computation of the Chern number as given by Eq.~\eqref{Cherneq1} is often challenging due to phase or gauge ambiguities~\cite{fukui2005chern}. Furthermore, it has to be done for each $\theta$ value that defines a different unit-cell size (see, for example, Refs.~\cite{rosa2019edge,riva2020edge}). Here, we take an alternative, and more generic, approach that produces the gap labels $C_g$ without direct computation of the band Chern numbers $C_n$, and for all $\theta$ values at once. Such approach relies on density of states computations based on the spectral variation with $\theta$, which has been developed using mathematical principles of K-theory in the context of periodic and aperiodic topological insulators~\cite{bellissard1992gap,prodan2016bulk}, and later extended to quasi-periodic acoustic/elastic metamaterials~\cite{apigo2019observation,ni2019observation,pal2019topological,gupta2020dynamics,xia2020topological,rosa2021exploring}. This approach has not yet been extended to continuous elastic periodic waveguides such as the beams studied here.

\subsubsection{Extraction of the Chern number by varying the modulation wavenumber}
\begin{figure*}[t!]
\centering
\includegraphics{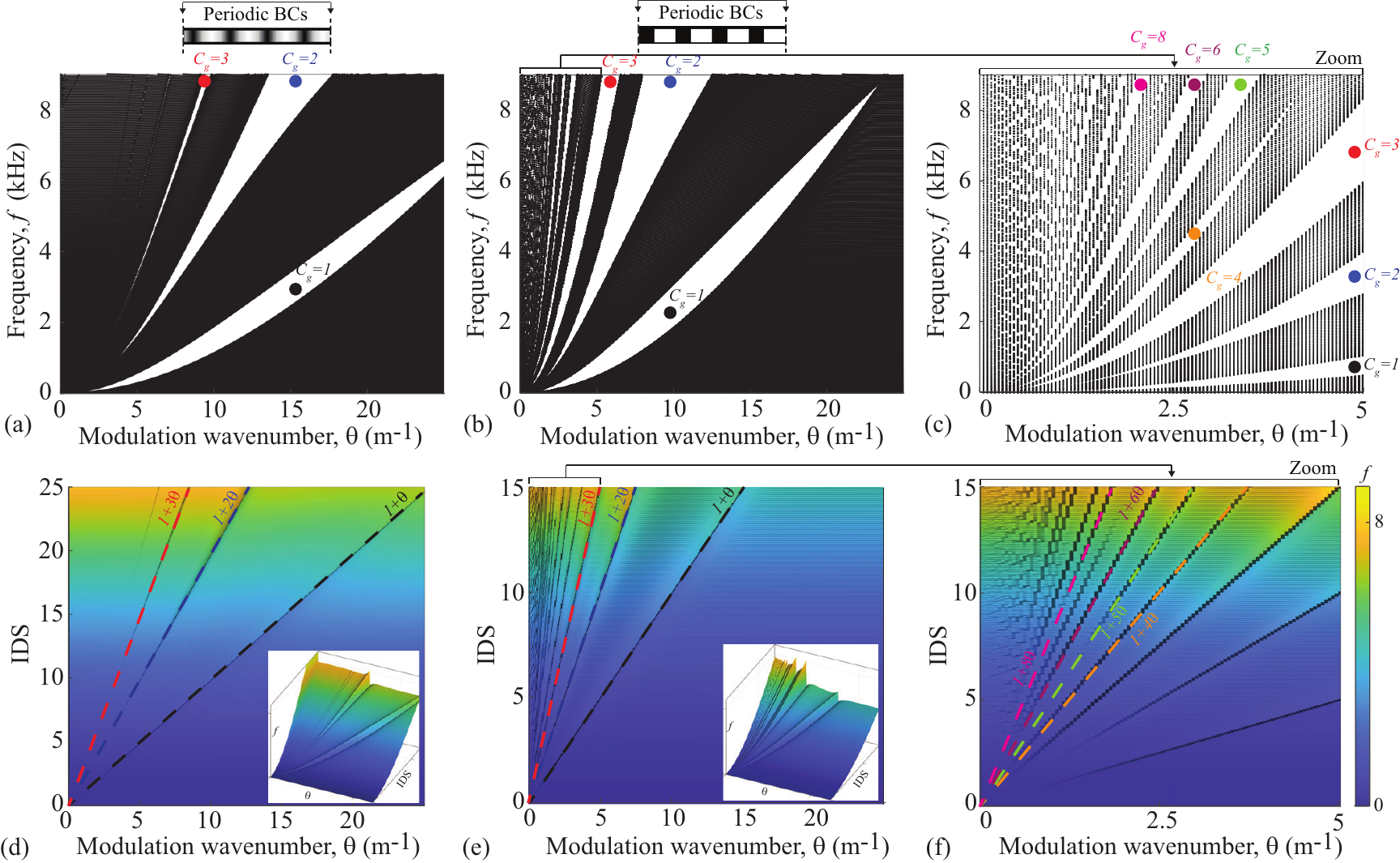}  
\caption{\label{fig:fig3}{Eigenfrequencies of finite beam with $L=100\bar{a}$ and PBCs for (a) sinusoidal and (b) step-wise modulation, with zoomed view in (c). Black dots represent eigenfrequencies while white areas denote band gaps. The corresponding IDS plots are displayed in the bottom panels (d-f), where selected fitted lines have colors corresponding to the gaps marked and labeled in (a-c).}}          
\end{figure*}

To begin, we investigate the variation of the beams' spectral properties as a function of the modulation wavelength $\theta$. The procedure relies on a large finite structure of fixed size $L=100\bar{a}$, and the computation of its eigenfrequencies under periodic boundary conditions (PBCs). The results are illustrated in Fig.~\ref{fig:fig3}(a,b) for the CM-PnC and SM-PnC configurations, where the eigenfrequencies are plotted as a function of $\theta$ as black dots. In the computation, the considered range of $\theta$ is discretized in intervals of $\Delta\theta =1/L$, i.e., $\theta_n=n/L$, such that each considered structure has an integer number $n$ of unit cells. By doing so, the resulting eigenfrequencies sample the Bloch dispersion bands defined for the considered $\theta$ value, and no frequencies are found inside the gaps due to the PBCs and the "perfect" periodicity emanating from an integer number of unit cells~\cite{pal2019topological}. The resulting spectrum provides a map for the location of the bands (black regions) and band gaps (white regions) as a function of $\theta$, and consequently of unit-cell length $a=1/\theta$. We note that SM-PnC produces a more complex spectrum (Fig.~\ref{fig:fig3}(b)) with a larger number of gaps when compared to CM-PnC (Fig.~\ref{fig:fig3}(a)), in particular for lower values of $\theta$ as illustrated in the zoomed view of Fig.~\ref{fig:fig3}(c).

The band-gap Chern numbers can be extracted by computing the Integrated Density of States (IDS) of the spectrum. It is defined as
\begin{equation}\label{IDSeq}
   \text{IDS} (\theta,f) = \lim_{L\to \infty} \dfrac{\sum_n [f_n\leq f ]}{L}, 
\end{equation}
where $[\cdot]$ denotes the Iverson Brackets, which provides a value of $1$ whenever the argument is true. In simple terms, for a given $\theta$ and frequency $f$, the IDS is the summation of all the eigenfrequencies below $f$, normalized by the structure size $L$. It theoretically converges as the structure size tends to infinity, but it is practically sufficient to consider large structures such as the one with $L=100\bar{a}$ considered in our investigation. The IDS is displayed for the CM-PnC medium in Fig.~\ref{fig:fig3}(d), and for the SM-PnC medium in Fig.~\ref{fig:fig3}(e) with a zoomed view for the lower $\theta$ range in (f). In this representation, the $z$-axis and the associated colormap represent frequency $f$ as a function of IDS and $\theta$. The insets in (d,e) illustrate the 3D views highlighting sharp discontinuities in the surface plot, which are visualized as straight lines in the top view colormaps. Each straight line is associated with a band gap and occurs since the IDS does not change inside the gap. Hence, a jump in frequency (color) occurs as the IDS changes from the last mode before the gap to the first mode right after the gap. According to the theory~\cite{apigo2019observation}, and confirmed by our findings, the variation of the IDS with $\theta$ inside the gaps identify straight lines expressed as
\begin{equation}
    \text{IDS}(f)=n_0+C_g\theta,
\end{equation}
with the gap Chern number $C_g$ corresponding to the slope. The lines of the most prominent gaps in Fig.~\ref{fig:fig3} are fitted and overlaid to the IDS plots, allowing the extraction of the Chern gap labels from the slopes as marked in the top panels, with different colors used to represent different gaps. These gap labels are defined generically for any $\theta$ value that defines the band gap, and are related to the truncation resonances as described in the following section. 

\subsection{Topological edge states and their control by phasons}
\label{ContTopStates}
\begin{figure*}[b!]
\centering
\includegraphics{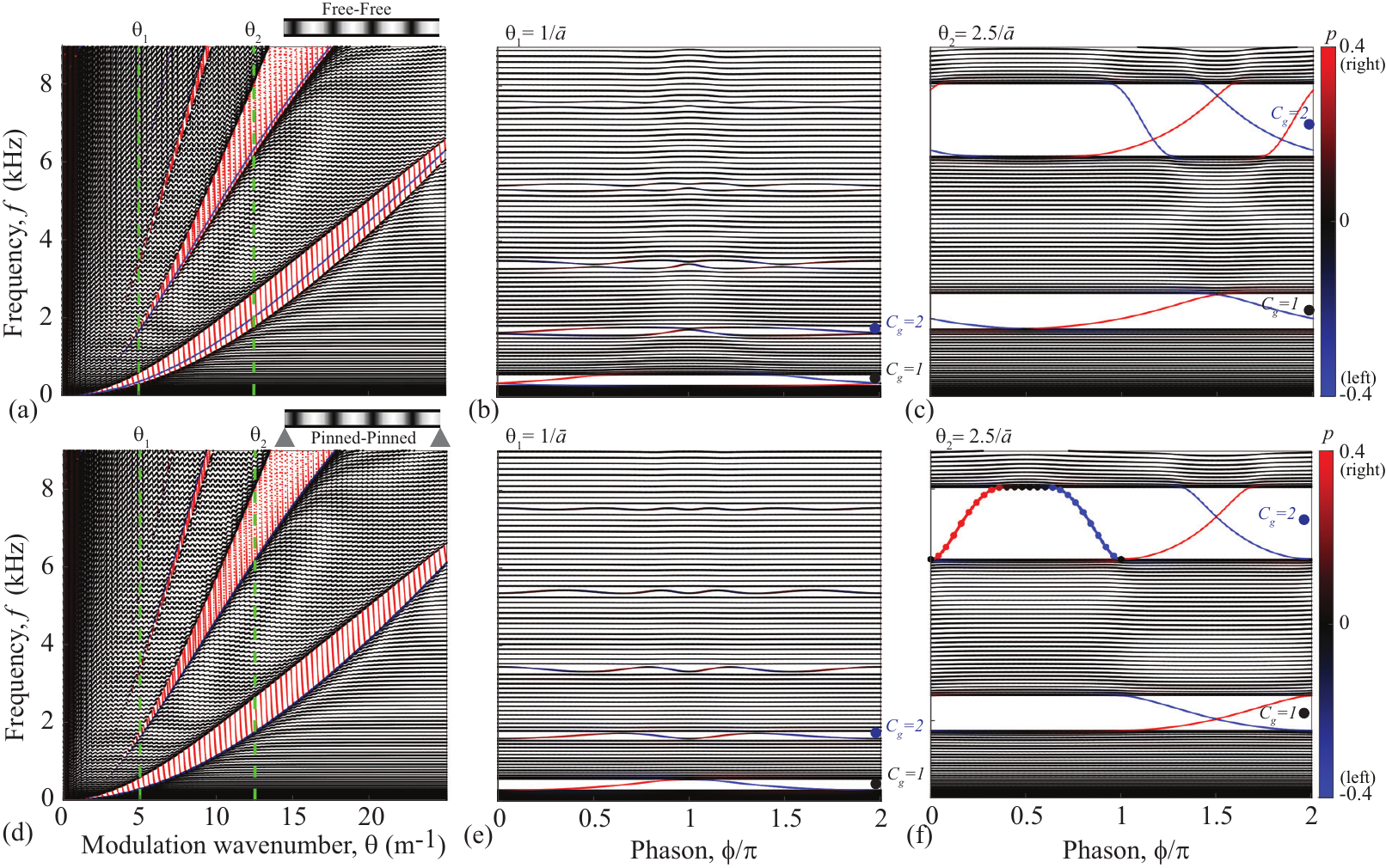}  
\caption{\label{fig:fig4}{Eigenfrequencies of finite CM-PnC structure with length $L=15\bar{a}$ and free-free (top) or pinned-pinned (bottom) BCs. The left panels (a,d) display the variation of the eigenfrequencies with $\theta$, while the middle (b,e) and right (c,f) panels display the variation with $\phi$ for the selected $\theta$ values highlighted as vertical dashed green lines in (a,d). The frequencies are color-coded according to the polarization $p$, and the gap labels $C_g$ are added for reference.}}          
\end{figure*}

\begin{figure*}[t!]
\centering
\includegraphics{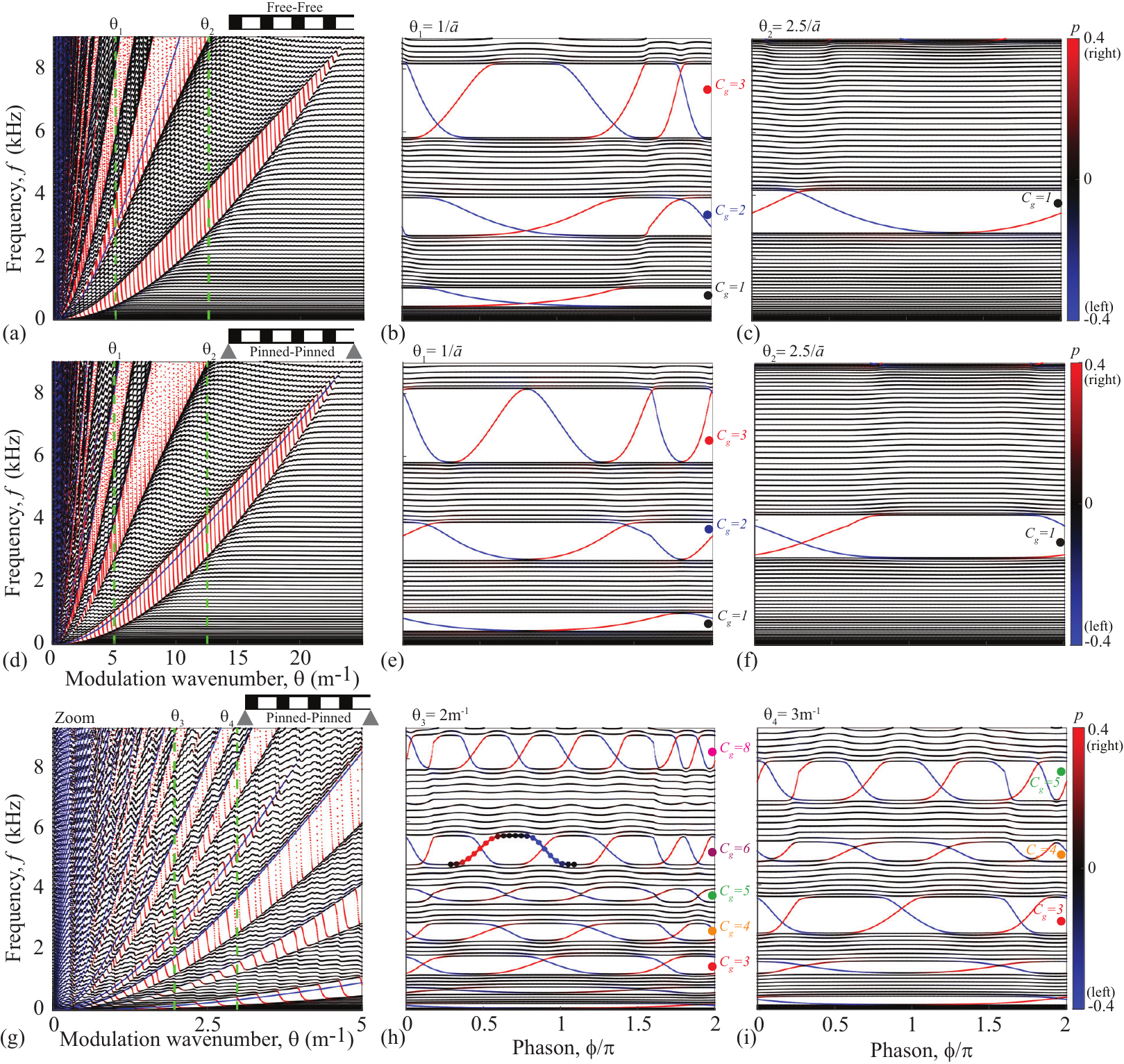} 
\caption{\label{fig:fig5}{Eigenfrequencies of the finite SM-PnC structure with length $L=15\bar{a}$ and free-free (top row) or pinned-pinned (middle row) BCs. The left panels (a,d) display the variation of the eigenfrequencies with $\theta$, while (g) displays a zoom of (d) in the low $\theta$ range. The middle (b,e,h) and right (c,f,i) panels display the variation with $\phi$ for the selected $\theta$ values highlighted as vertical dashed green lines in (a,d,g). The frequencies are color-coded according to the polarization $p$, and the gap labels $C_g$ are added for reference.}}          
\end{figure*}

The non-zero Chern gap labels indicate the presence of in-gap edge states existing for structures with truncated boundaries, i.e., the truncation resonances. Their properties are illustrated in Figs.~\ref{fig:fig4} and ~\ref{fig:fig5} for the CM-PnC and SM-PnC configurations, respectively. The figures display the frequencies of a finite structure of fixed length $L=15\bar{a}$ as a function of modulation wavenumber $\theta$ and phase $\phi$, for different BCs such as free-free and pinned-pinned. The frequencies are color-coded according to a localization factor $p$ to identify modes localized at the boundaries, which is defined as
\begin{equation}\label{eq:locfactor}
    p=\dfrac{\int_{\mathcal{L}_r} |w| dx - \int_{\mathcal{L}_l} |w| dx}{\int_{\mathcal{L}} |w| dx},
\end{equation}
where $\mathcal{L}$ denotes the domain of the beam, and $\mathcal{L}_r$ and $\mathcal{L}_l$ correspond to a smaller portion of length $0.15L$ at the right and left boundaries, respectively. With this definition, positive (red) and negative (blue) $p$ values indicate modes localized at the right and left boundary, respectively, while values that are close to zero (black) indicate non-localized bulk modes. 

The left panels in Figs.~\ref{fig:fig4} and ~\ref{fig:fig5} display the eigenfrequencies of the finite beam as a function of $\theta$, for different BCs as illustrated by the schematics. The spectra are overall similar to the bulk spectra exhibited in Fig.~\ref{fig:fig3}, with black regions also defining the bulk bands, but with additional modes appearing inside the band gaps. These modes are the topological edge states, corresponding to the truncation resonances which are localized at one of the boundaries of the beam. The modes localized at the right boundary (red) traverse the band gaps multiple times as they migrate from the band above to the band below their respective gaps. Although not the focus of the present investigation, this behavior stems from the positive gap labels $C_g>0$ and can be explained by density of states arguments~\cite{pal2019topological}. Furthermore, the modes localized at the left boundary (blue) do not migrate between bands and instead remain inside the gap for the considered range of $\theta$. The different behavior between left- and right-localized modes occur due to the way the finite structure is constructed, where the change in $\theta$ produces a qualitative change at the right boundary (the modulation is truncated at different places for different $\theta$), but not of the left boundary (the modulation is always truncated at the same place). 

The gap label $C_g$ dictates the number of left- and right-localized edge modes that span the band gap as the phason $\phi$ varies within an interval of $2\pi$, for a fixed $\theta$ value. This is illustrated for selected $\theta$ values (marked as vertical dashed green lines) in the middle and right panels of Figs.~\ref{fig:fig4} and \ref{fig:fig5}, which display the variation of the eigenfrequencies with the phason $\phi$. As previously mentioned, variations of $\phi$ do not affect the frequencies of the dispersion bands, and therefore the boundaries of the band gaps (material property) remain unchanged with $\phi$. However, the phason influences how both boundaries of a finite structure are truncated (Fig.~\ref{fig:fig1}), and its variation causes the eigenfrequency branches of the truncation resonances to traverse the gaps. The first selected value $\theta_1=1/\bar{a}$ corresponds to the modulation wavelength for the reference unit-cell size $\bar{a}$. In the CS-PnC case (Figs.~\ref{fig:fig4}(b,e)), this unit-cell size produces two small gaps with Chern labels $C_g=1$ and $C_g=2$, which were extracted from the procedure in Fig.~\ref{fig:fig3}. For both types of BCs (free-free in (b) and pinned-pinned in (e)), one left- and one right-localized edge state traverse the first gap, and two edge states traverse the second gap, as the phason $\phi$ varies from $0$ to $2\pi$. In the SM-PnC case (Figs.~\ref{fig:fig5}(b,e)), the choice $\theta_1=1/\bar{a}$ corresponds to the case investigated in the experimental section of this paper (see Section~\ref{Sec:Experiments}), which produces three band-gaps with $C_g$ values ranging from $1$ to $3$. Regardless of the type of boundary condition, the number of left- and right-localized edge modes spanning the band gaps is equal to the corresponding Chern gap label. In addition, the gap label sign is related to the direction the edge modes cross the gap~\cite{prodan2016bulk}. A positive $C_g>0$ indicates that $|C_g|$ left-localized branches will cross the gap from the lower band to the upper band, and an equal number of right-localized states will cross from the upper band to the lower band. Although no examples are found in this paper, a negative sign $|C_g|<0$ produces transitions in opposite directions~\cite{rosa2019edge}. Also note that the eigenfrequencies have a periodic behavior with $\phi$, and are actually continuous at $\phi=0=2\pi$. Therefore a few branches of the truncation resonances traverse the gap through that point; for example, see the second right-localized mode in the second gap of Fig.~\ref{fig:fig5}(e). Indeed, the phason variable $\phi$ defines a continuous ring, with no start or ending point, with the beginning and end at $\phi=0$ and $\phi=2\pi$, respectively, being arbitrary choices for the plots.  

\begin{figure*}[t!]
\centering
\includegraphics{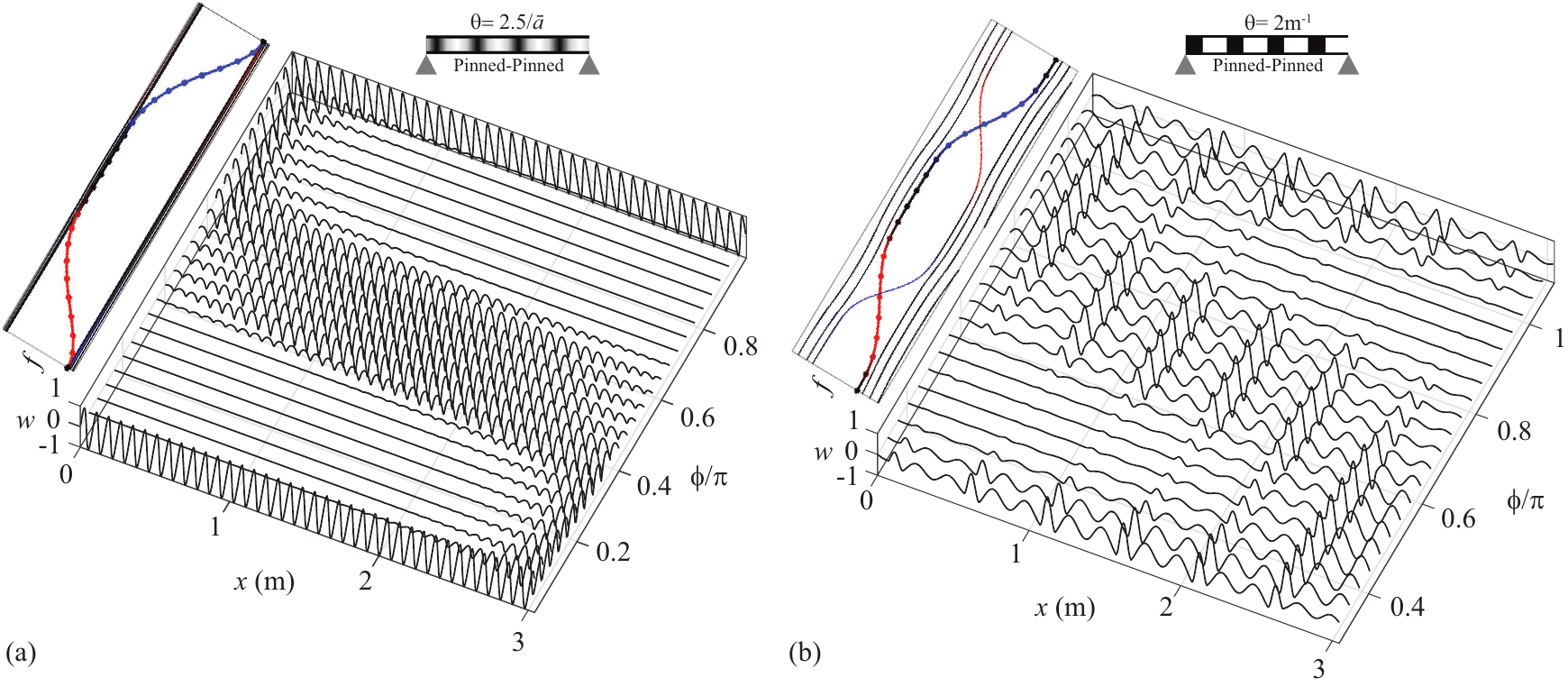}  
\caption{\label{fig:fig6}{Examples of mode shape transitions as a function of phason $\phi$ for the (a) CM-PnC and (b) SM-PnC structures, corresponding to the branches highlighted in Fig.~\ref{fig:fig4}(f) and Fig.~\ref{fig:fig5}(h), respectively.}}   \end{figure*}

Other examples are shown to demonstrate the generality of the approach and give more insights into the behavior of the edge states. The case of $\theta_2=2.5/\bar{a}$ (panels (c,f) in Figs.~\ref{fig:fig4} and \ref{fig:fig5}) corresponds to a unit-cell size 2.5 times smaller than the reference $\bar{a}$, and therefore the finite length $L=15\bar{a}$ now comprises $37.5$ unit cells. Even without an integer number of unit cells, the number of edge sates inside each gap matches the corresponding gap labels, for both CS-PnC and SM-PnC, and both types of BCs considered. In fact, this behavior is general and holds for any arbitrary $\theta$ value. The last row in Fig.~\ref{fig:fig5} focuses on the lower $\theta$ range, where the SM-PnC features additional gaps with higher Chern gap labels. The examples $\theta_3=2$ m$^{-1}$ and $\theta_4=3$ m$^{-1}$ correspond to unit cell sizes of $0.5$m and $0.33$m respectively, and form finite structures with 6.09 and 9.135 unit cells for the fixed length $L=15\bar{a}$. They feature gap labels as high as $C_g=8$, and the behavior of the edge states spanning the gaps with $\phi$ is in agreement with the extracted gap labels, again even without an integer number of unit cells. Among many edge states, two transitions experienced by the modes as a function of $\phi$ are highlighted by thicker lines and dots in Fig.~\ref{fig:fig4}(f) and in Fig.~\ref{fig:fig5}(h), and have their mode shape variation displayed in Figs.~\ref{fig:fig6}(a,b) respectively. These examples illustrate a transition between a right- and left-localized mode that occurs as a function of $\phi$, with an intermediate state as a non-localized bulk mode when the eigenfrequency branch tangentially approaches the boundary of the gap. This type of transitions have been exploited for topological pumping applications, where the phason $\phi$ is varied along an additional spatial~\cite{rosa2019edge,riva2020edge} or temporal~\cite{grinberg2020robust,chen2019mechanical,xia2021experimental}  dimension to induce a migration of localized modes between two boundaries. 

These results reveal that the truncation resonances are in fact topological edge states that traverse the band gaps for variations of the phason $\phi$. The number of truncation resonances that traverse a gap is equal to the corresponding gap label $C_g$. This holds true for any set of BCs, although the particular shape of the branches of the edge states as they traverse the gap may be different. In addition, while the number of in-gap resonances can be predicted, one cannot guarantee the existence of truncation resonances for a particular phason value $\phi$, but only that $|C_g|$ branches will traverse the gap when $\phi$ varies in an interval of $2\pi$. For example, the finite structure considered in Fig.~\ref{fig:fig2}(a) correspond to a phason value $\phi=0.2\pi$, which intersects both the right- and left-localized edge state branches of Fig.~\ref{fig:fig4}(b), and therefore one resonance localized at each boundary is found in this case. In contrast, for a phason value $\phi=\pi$, the same gap in Fig.~\ref{fig:fig4}(b) does not exhibit any edge states, and therefore no truncation resonances would be found. Similarly, the modes I and II in Fig.~\ref{fig:fig2}(b) are intersections of the left- and right-localized edge state branches in the first and third gap of Fig.~\ref{fig:fig5}(b), respectively, for $\phi=0.4\pi$, while other phason choices would define different truncation resonances or the their absence. Therefore, to better understand the behavior of the truncation resonances one needs to consider the entire family of structures defined for variations of $\phi$, instead of separately considering particular cases. 

\subsubsection{Boundary phasons}
\label{BoundPhason}
\begin{figure*}[t!]
\centering
\includegraphics{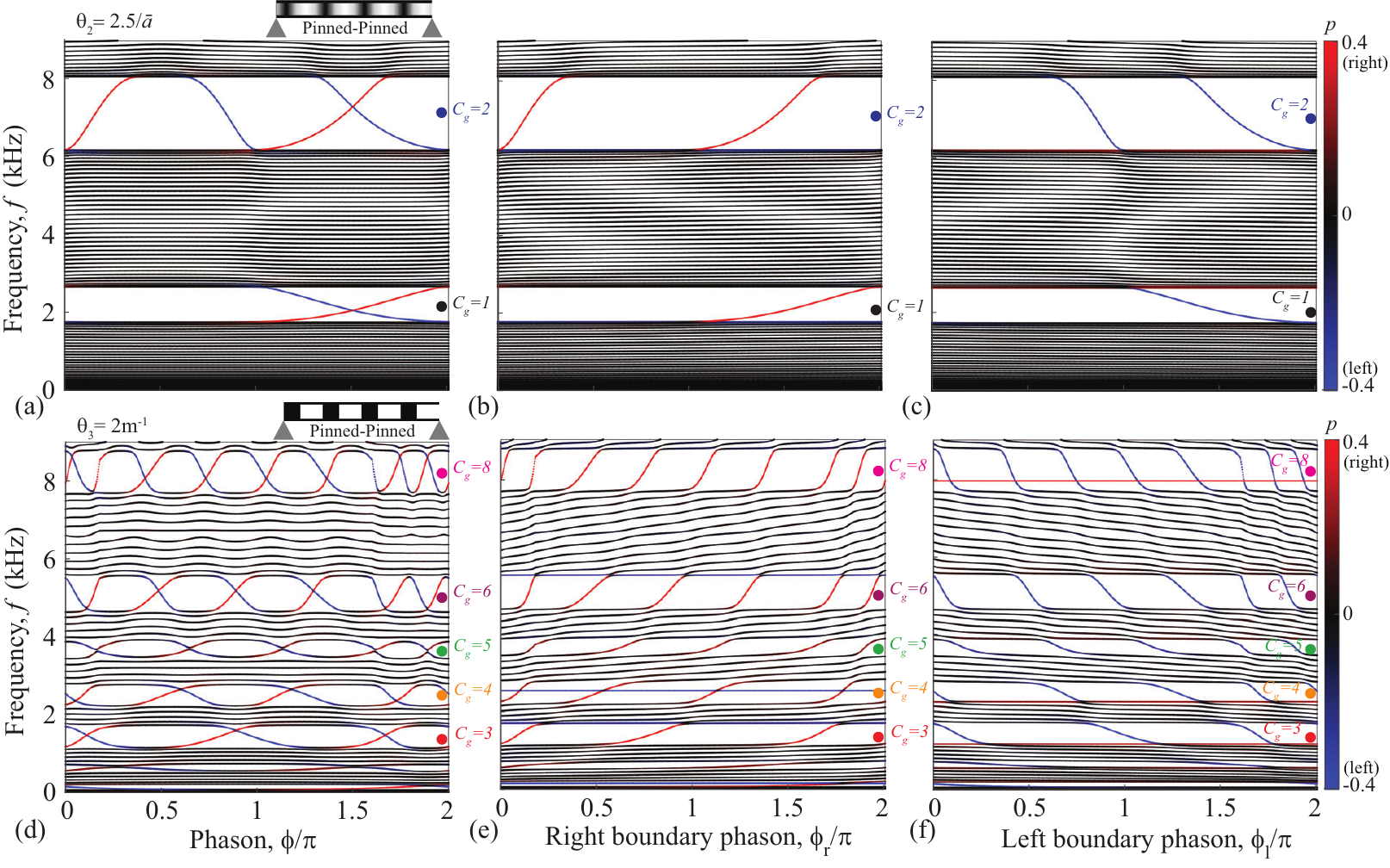}  
\caption{\label{fig:fig7}{Eigenfrequency variation as a function of phason $\phi$ (a,d), right-boundary phason $\phi_r$ (b,e) and left-boundary phason $\phi_l$ (c,f) for finite beam with $L=15\bar{a}$ and pinned-pinned BCs. The top row consists of a CM-PnC structure with $\theta_2=2.5/\bar{a}$ while the bottom row consists of a SM-PnC structure with $\theta_3=2$ m$^{-1}$.}}          
\end{figure*}

As described, the phason $\phi$ simultaneously modifies the properties of both boundaries of a finite structure (Fig.~\ref{fig:fig1}), and therefore influence the truncation resonances localized at both boundaries. A higher degree of control over the truncation resonances is achieved by using the right- and left-boundary phasons introduced in Fig.~\ref{fig:fig1}, which modify only one boundary at a time. This is equivalent to adding a tuning layer at one end of the structure as done in Refs.~\cite{Hussein_JSV_2007,DavisIMECE2011}. The effect of boundary phasons is demonstrated in Fig.~\ref{fig:fig7}, which repeats the eigenfrequency variation with $\phi$ of Fig.~\ref{fig:fig3}(f) and Fig.~\ref{fig:fig4}(d) in the left panels, and compares them to the the variation as a function of right-boundary phason $\phi_r$ and left-boundary phason $\phi_l$ displayed in the middle and right panels, respectively. The plots clearly show evidence of how the boundary phason only causes the edge states localized at the corresponding boundary to traverse the gap, while the superimposed effect of both boundary phasons lead to the effect caused by the phason $\phi$.  Indeed, as $\phi_r$ varies (Figs.~\ref{fig:fig7}(b,e)), only the right-localized modes traverse the gaps, producing the same branches as the ones in Figs.~\ref{fig:fig7}(a,d). Any left-localized modes that were defined for $\phi=0$ (the starting point) appear as roughly flat bands inside the gap, since the left boundary is not changing with $\phi_r$. A similar effect is observed for the variation with $\phi_l$ in Figs.~\ref{fig:fig7}(c,f). For a structure that has a sufficient number of unit cells (i.e., has reached convergence as described Section~\ref{EffectNC} to follow), the right- and left-localized edge states form a set of decoupled chiral bands~\cite{prodan2016bulk}, the number of which corresponds to the gap label magnitude $|C_g|$ and whose slopes are associated with the gap label sign. \\

\subsection{Effect of number of unit cells on frequency convergence of topological truncation resonances}
\label{EffectNC}

Next, we investigate the effect of the number of unit cells on the behavior of the truncation resonances. As shown earlier, truncation modes exhibit an exponential decay away from the boundary since their frequency lies inside a band gap, and therefore correspond to a complex wave number. For structures with a large number of unit cells, the in-gap truncation modes are only mildly affected by further addition of unit cells since their displacement tend to zero away from the boundary. In that scenario, a further increase in number of unit cells will produce a larger number of bulk modes, while the branches of the edge states spanning the band gaps with $\phi$ will remain the same. However, for structures with a small number of unit cells, the truncation resonances are more likely to be  influenced by the opposing edge and by other effects such as mode coupling and veering with bulk modes or another edge state. 

\begin{figure*}[b!]
\centering
\includegraphics{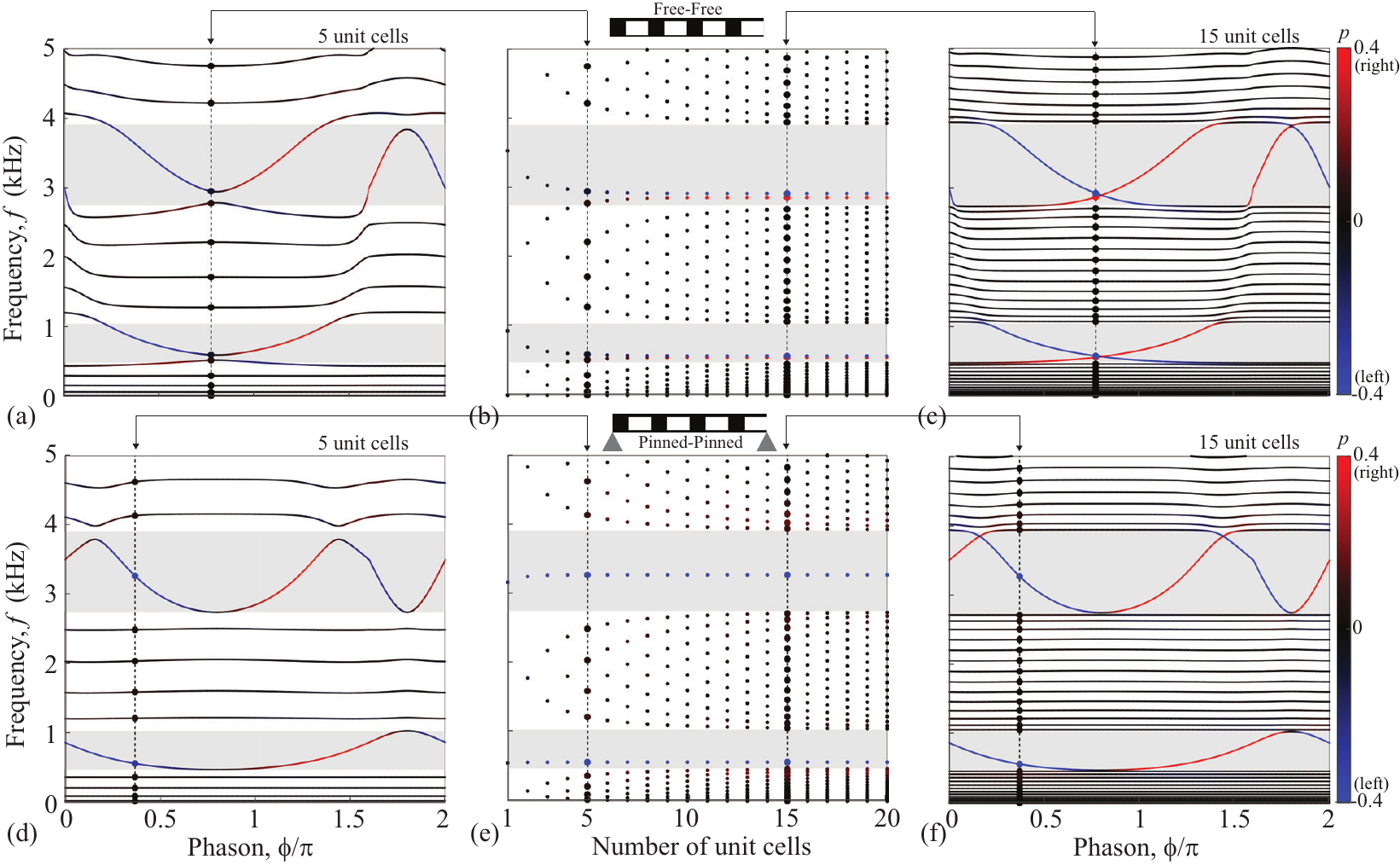}
\caption{\label{fig:fig8}{Eigenfrequency variation with $\phi$ for structure with $\theta_1=1/\bar{a}$ comprising 5 cells (a,d) and 15 cells (c,f). The middle panels (b,e) show the variation with the number of unit cells for the fixed phason values highlighted as vertical dashed lines in the other panels. Top and bottom rows correspond to free-free and pinned-pinned BCs, respectively. Band-gap frequency ranges are shaded grey.}}          
\end{figure*}

This behavior and the convergence with the number of unit cells is elucidated by the results of Fig.~\ref{fig:fig8}. The SM-PnC structure with $\theta_1=1/\bar{a}$ is chosen to exemplify these features, with the first and second row corresponding to free-free and pinned-pinned BCs, respectively. The panels (a,d) display the variation of the eigenfrequencies with $\phi$ for a structure with 5 unit cells, while the right panels (c,f) correspond to a larger structure comprising 15 cells. In the middle panels (b,e), the variation of the frequencies with the number of unit cells is displayed for the fixed phason value highlighted by the vertical dashed-line intersections in the other panels. Overall, the number of bulk modes increase with the number of unit cells as expected, and the edge state branches traversing the gaps are similar but exhibit small differences. These differences are amplified for phason values that are close to mode couplings as illustrated in the top row. At the selected phason value, there is a strongly coupled avoided crossing between the right- and left-localized edge states for the case with 5 unit cells shown in (a), and therefore the eigenfrequencies defined for that phason value are more separated when compared to the structure 
shown in (c) with 15 cells and without the avoided crossing. Therefore, the frequencies of the edge states for this phason value vary as a function of the number of unit cells and converge to a fixed value at approximately 10 unit cells as illustrated in Fig.~\ref{fig:fig8}(b). In contrast, in the case of the bottom row with pinned-pinned BCs, the chosen phason value intersects the edge state mode and an adjacent mode that is well isolated, and therefore the truncation frequency converges quicker at around four unit cells. These results illustrate that while convergence is always achieved, the required number of unit cells may vary between different structures depending on the BCs and the presence of coupling effects at the phason value of interest. 


\subsection{Topological truncation resonance versus non-topological defect resonance}
Truncation resonances, with their topological character, are not the only type of resonances that appear due to truncation or breakage of symmetry in a periodic medium. Another type of resonance, that is also of localized nature, is that associated with defect modes~\cite{Sigalas1997,Torres1999,Khelif2003}. Under the developed framework, the band gaps characterized by non-zero Chern labels are guaranteed to support $|C_g|$ truncation resonances spanning the gaps as a function of phason or boundary phason parameters. Although we do not present an example in this paper, in some cases a band gap may be characterized by $C_g=0$, which is referred to as a topologically trivial band gap. In this case, the presence of in-gap resonances is not guaranteed, although they may appear. Since there is no topological explanation or origin to their appearance, these truncation resonances are usually categorized as defect modes. One example can be found in reference~\cite{xia2020topological}, where a central trivial gap with $C_g=0$ does not exhibit in-gap resonances under pinned-pinned BCs (Fig. 2a), but exhibits truncation resonances under clamped-free BCs (Fig. 3a). Note that the truncation resonances in this second case do not traverse the band gap, which is a key feature expected from topological modes as we highlight in this work.

\begin{figure*}[t!]
\centering
\includegraphics{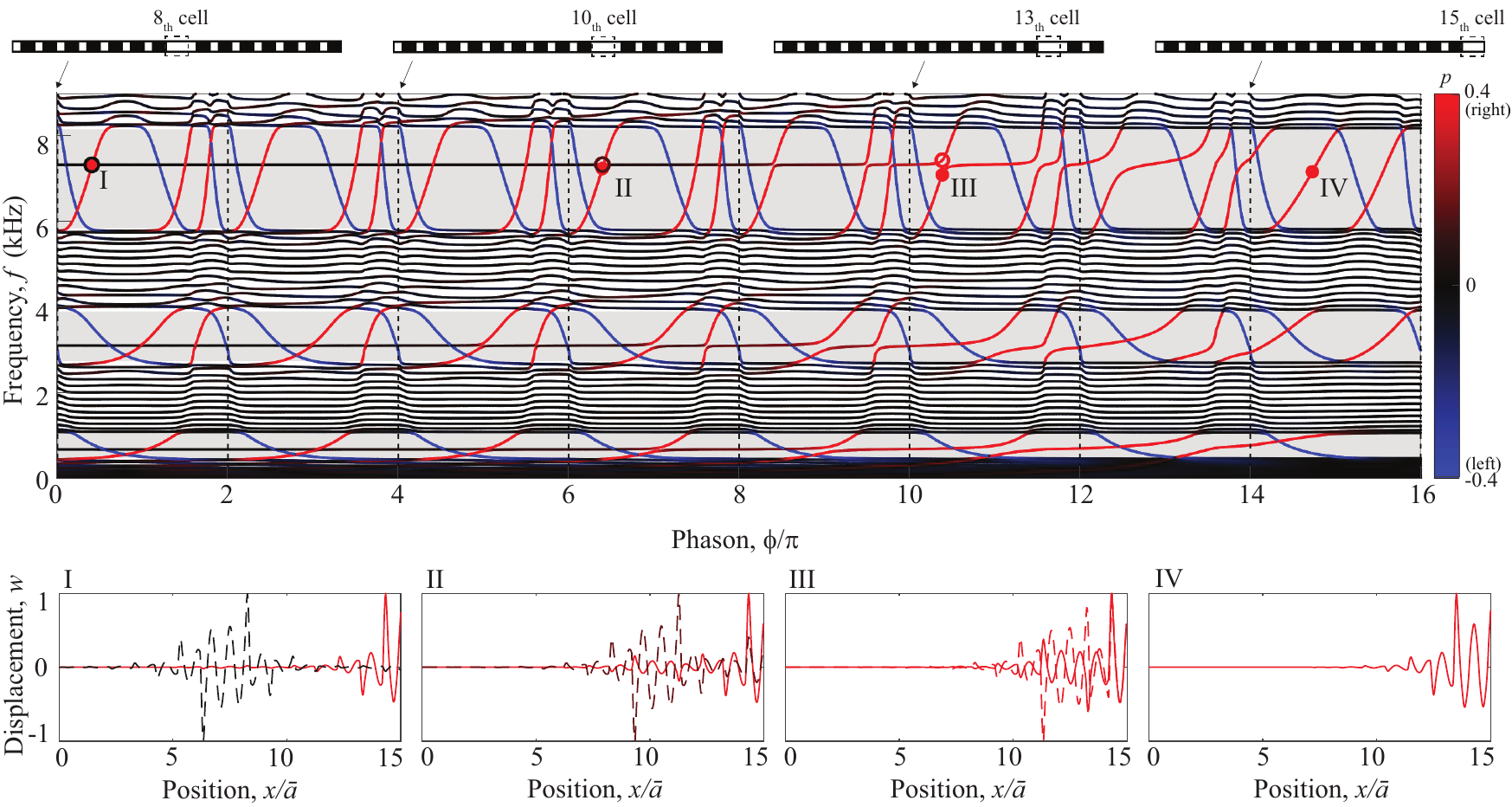} 
\caption{\label{fig:fig9}{Eigenfrequencies as a function of phason $\phi$ for a finite SM-PnC structure with $\theta_1=1/\bar{a}$, $L=15\bar{a}$ and a defected unit cell. The location of the defect changes by one unit-cell increments with every change of $2\pi$ in $\phi$, as marked by the vertical dashed lines and illustrated in the top schematics. Band-gap frequency ranges are shaded grey. Selected mode shapes are displayed in the bottom panels, whose colors correspond to the polarization of the mode, with dashed and solid lines representing the mode with open and closed circle markers, respectively.}}   \end{figure*}

We here illustrate another important scenario where a physical defect is introduced to a finite structure in order to create an in-gap resonance, although in this case a non-topological resonance as we will show. As an example, we consider a finite SM-PnC structure comprising 15 unit cells with $\theta_1=1/\bar{a}$, and introduce a defect initially located at the 8th unit cell by "skipping" the ABS portion within this unit cell, making it entirely out of aluminum. The results displayed in Fig.~\ref{fig:fig9} show the variation of the eigenfrequencies with $\phi$, with the defect unit cell highlighted in the schematics at the top and identified by the larger white segment, which represents aluminum. As the phason varies, material is added to the left boundary and removed from the right boundary (Fig.~\ref{fig:fig1}), which causes the defect to continuously drift towards the right boundary. The defect moves by one unit cell with every change in $2\pi$; these increments are marked by the vertical dashed lines in the figure. After a change in phason of $14\pi$, the defect is at the last unit cell, and finally for $16\pi$ it exists the structure and a perfect periodic domain is restored. In a defect-free structure, the variation of the eigenfrequencies with $\phi$ is trivially periodic in intervals of $2\pi$. With the inclusion of the defect, additional modes are found inside the gaps and co-exist with the truncation resonances. The interplay between the in-gap defect mode and the truncation resonances is highlighted by the selected mode shapes displayed in the bottom panels. In the initial configuration, the in-gap defect resonance is localized at the center (8th unit cell) of the structure and is completely decoupled from the truncation resonances, as evidenced by the plots in stage I. As the phason varies, the trajectory of the defect modes remain almost flat inside the gaps, in sharp contrast to the behavior of the topological states which transverse the gaps. Indeed, the truncation resonances exhibit the expected periodic behavior as their branches traverse the gaps in a pattern that repeats periodically in intervals of $2\pi$. However, as the defect physical position approaches the right boundary, the in-gap defect modes progressively couple with the truncation resonances localized at the right boundary, this is seen in all three gaps viewed in the figure. Focusing on the third band gap as an example, the frequency curves in stage II exhibit a weak coupling, while in stage III a larger coupling is observed causing an avoided crossing with relatively strong repulsion between the defect and truncation resonances. As the defect moves within the last unit cells (13th-15th), it slowly transforms to capture, itself, the characteristics of a truncation resonance localized at the right boundary, with a mode shape example displayed for stage IV. At this last stage, the branches of the right-localized truncation resonances are very different from the periodic pattern of the perfect periodic structure, since they are created by a truncation near a defect.

These results highlight key differences between the truncation resonances and defect modes. The defect mode defines a flat branch inside the gap as a function of $\phi$, until it starts to couple with the topological truncation resonances$-$which happens as the position of the defects nears the boundary. It is interesting to note that when the coupling takes place, the shape of the coupled truncation resonance branch changes as it traverses the gap. However, the counting principle given by the gap Chern labels is still valid. This can be verified as in every interval of $\phi=2\pi$, there is a net number of 1, 2 and 3 right-localized modes transversing the first, second, and third gap, respectively. Therefore, the truncation resonances retain this key topological property even with the interference of a defect at the boundary. We should also stress that the topological classification of an in-gap mode is always relative to a given set of parameters. The defect mode introduced here is non-topological in the context of the phason degree of freedom, which causes it to remain confinded inside the gap as a flat band. However in some cases this type of defect mode might find a topological classification under a different set of parameters and analysis framework~\cite{miniaci2021spectral}. 

\section{Experimental investigation of modulated phononic crystal beams}\label{Sec:Experiments}

\subsection{Experimental set-up and measurements}
\label{ExpSet}
\begin{figure*}[b!]
\centering
\includegraphics{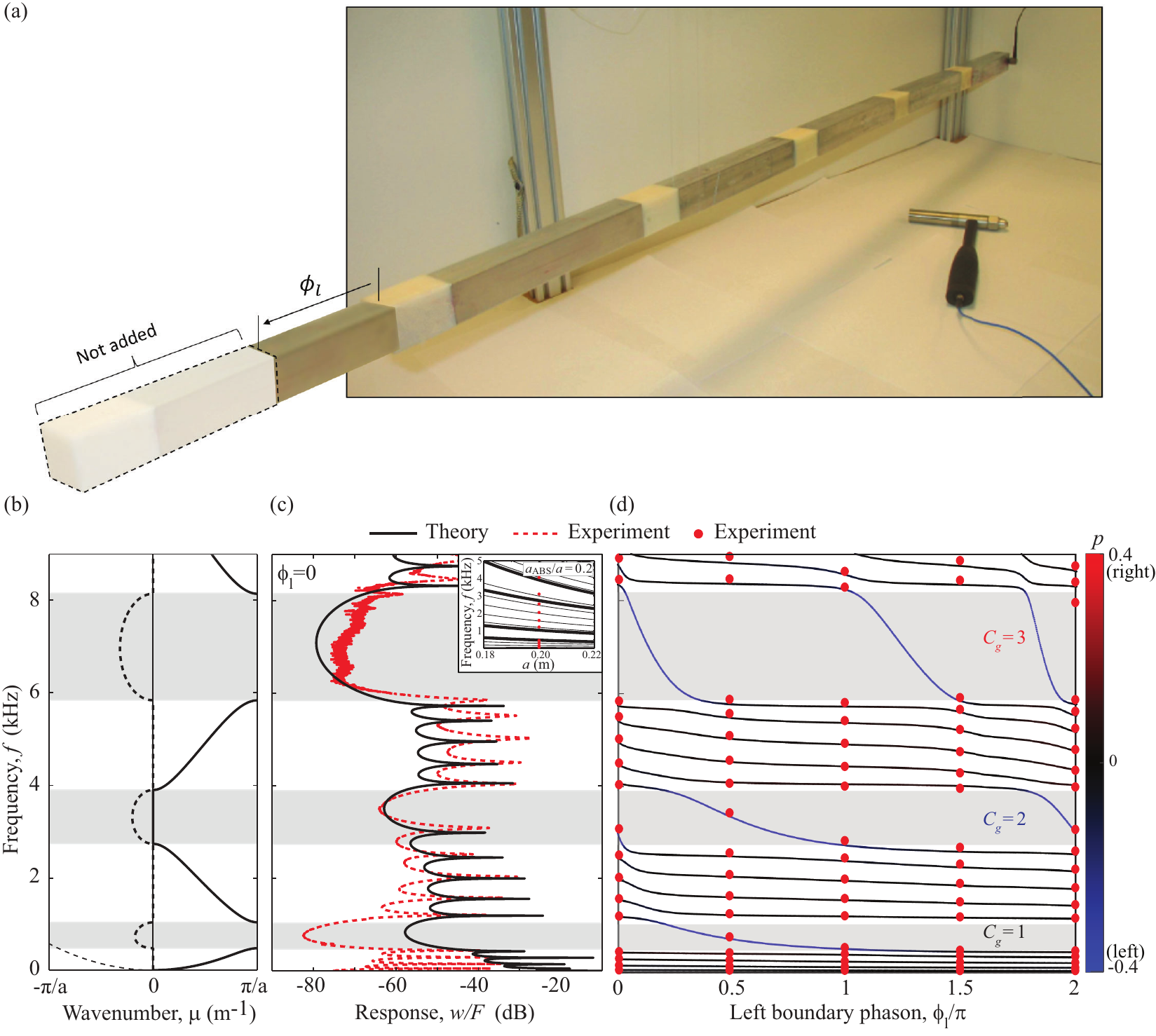}  
\caption{\label{fig:fig10}{Experimental validation: (a) Photograph of the experimental setup showing a 5-unit-cell SM-PnC beam structure consisting of layers of Aluminum and ABS polymer with a ABS volume fraction of $20\%$ and $\bar{a} = 203~\text{mm}$. The structure was excited on the far left side (on the first ABS polymer layer) with a force hammer and measured with an accelerometer on the other far end. (b) Frequency band diagram of the infinite (material) constituent of the SM-PnC beam and (c) corresponding FRF response of the finite structure.  Inset: Resonance frequency (thin solid lines, theory; dots, experiment) versus unit-cell length $a$ for the 5-unit-cell periodic beam structure. (d) Corresponding resonance frequency (solid lines, theory; dots, experiment) versus left boundary phase (i.e., length of a tuning layer attached at the far left end). At $\phi_{\rm l}=0.4 \pi$, the tuning layer transitions from ABS to Al. At $\phi_{\rm l}=2 \pi$, the tuning layer is a full regular unit cell and the total structure is rendered a 6-unit-cell structure. In (a), the solid lines represent propagation modes, and the dashed lines represent attenuation modes. Band-gap frequency ranges are shaded grey.}}          
\end{figure*}

For the experimental investigation, we focus on the SM-PnC beam structure, again composed of alternating layers of Al and ABS with a ratio of layer lengths of 4:1 (Al:ABS) for the baseline unit-cell configuration. The unit-cell length and cross-sectional area are selected as $\bar{a} = 203~\text{mm}$ and $A = 645~{\text{mm}^2}$, except in Section~\ref{CellVolF} where the unit-cell length is varied. The values of these geometric parameters are chosen to allow for the generation of several band gaps below 9 kHz for practical reasons; however, all conclusions are scale invariant and hence applicable to periodic structures that are orders of magnitude smaller in size (with the limit that they are appropriately represented by continuous models). In this section, we show additional FE results for direct comparison with the experiments, where we use the same FE model details as in Section~\ref{Sec:Theory} with specifically 100 finite elements being used per unit-cell. For our experimental set-up, a set of Al and ABS solid blocks were fabricated and connected to each other by an adhesive to form the periodic structure. The test articles were suspended using thin nylon wires to simulate free-free BCs as depicted in Fig.~\ref{fig:fig10}(a).

First we show the complex band structure of the unit cell, which is shown in Fig.~\ref{fig:fig10}(b)$-$the real part of which is identical to Fig.~\ref{fig:fig2}(b). This calculation shows that three relatively large band gaps exist between 0 and 9 kHz. Figure.~\ref{fig:fig10}(c) shows a corresponding FRF obtained theoretically (solid line) and experimentally (dashed line) for a 5-unit-cell version of the structure, in which the “input” force excitation and the “output” displacement evaluation are at the extreme ends. For the experimental results, the test article was excited at the tip of the structure using a force hammer. The impulse forcing data $F$ from the force hammer was used in conjunction with the response data $U$ obtained by a sensing accelerometer connected at the other end of the structure, to generate the receptance $U/F$ over the frequency range 0-9 KHz. The amplitude of the experimental response was calibrated to match the average of all theoretical data points over the 0-9 KHz frequency range. An excellent correlation is observed between the theoretical and experimental FRF curves. It can be seen, however, that the correlation generally degrades at higher frequencies along with an increasing level of noise. This is due to the difficulty of stimulating high frequencies with a force hammer as well as the reduced resolution when using a constant sampling rate over all frequencies. 

\subsection{Effects of modulation wavenumber, boundary phasons, and number of unit cells by experiment}
\label{ModBoundNum}
In Fig.~\ref{fig:fig5}(a), we have shown the effect of the modulation wavenumber (i.e., unit-cell length) on the locations of the truncation resonances. Here we repeat our computational investigation focusing on the range $0.18\le a\le 0.22$ m and overlay the data of the experimental case of $a=0.2$ m $(\theta=5)$. The results, which are shown in the inset of Fig.~\ref{fig:fig10}(c), indicate very good agreement between theory and experiments. Another approach that keeps the unit-cell geometric configuration intact is the addition of a single tuning layer (or a partial unit-cell) at the end of the finite periodic structure~\cite{Hussein_JSV_2007,DavisIMECE2011}, as demonstrated in Section\ref{BoundPhason}. As illustrated in Fig.~\ref{fig:fig1}, the addition of a tuning layer corresponds to the application of a boundary phason $\phi_l$. The material and geometrical  configuration of the tuning layer should be chosen such that it would generally form a physically cropped unit-cell, i.e., it would form a partial unit-cell when its length is less than $a$ and a full unit-cell when its length is $a$.~Figure~\ref{fig:fig10}(d) displays a plot of the resonant frequencies as a function of the length of the tuning layer, denoted by $l_{\text{TL}}$ and ranging from  $l_{\text{TL}}$ = 0 ($\phi_l=0$, 5 unit-cells) to $l_{\text{TL}}=a$ ($\phi_l=2\pi$, 6 unit-cells) for the same baseline design of Fig.\ref{fig:fig10}(a)$-$this corresponds partially to the results shown in Fig.~\ref{fig:fig5}(b) but now with the addition of experimental data points. With the addition of a tuning layer, band-gap resonances rapidly traverse the band gaps. However, once they reach the band-gap boundaries they behave like regular structural resonances (bulk modes) with slower levels of variation as a function of $l_{\text{TL}}$.  

\begin{figure*}[t!]
\centering
\includegraphics{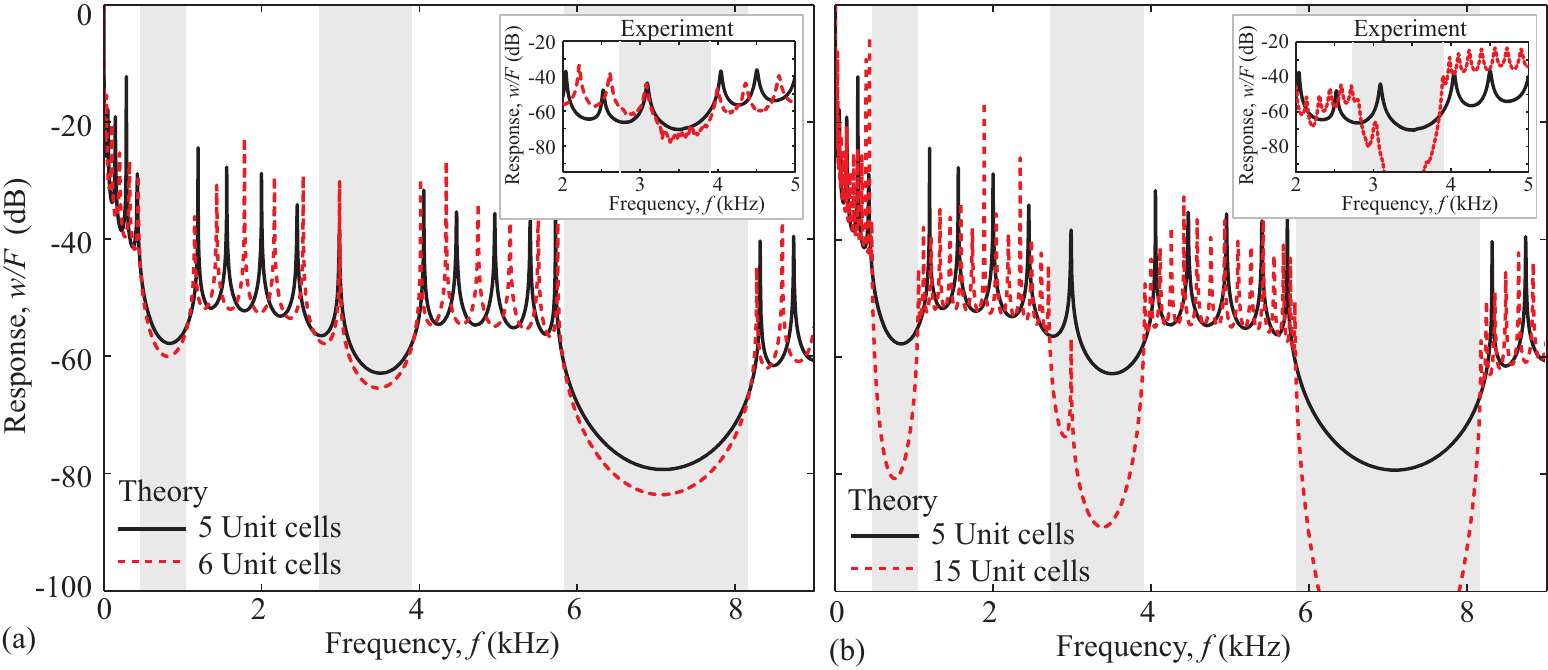}  
\caption{\label{fig:fig11}{Frequency response function comparison for the finite SM-PnC beam structure with different number of unit cells. The results show a truncation resonance in the second band gap. Compared to the baseline case of 5 unit cells, the tructioan resonance is observed to experience negligible shift in frequency for (a) a 6 unit-cell structure and (b) a 15 unit-cell structure. Strong spatial attenuation in displacement amplitude across the structure is observed as the number of unit cells is increased. These results are for the same unit-cell configuration considered in Fig.~\ref{fig:fig10}. Band-gap frequency ranges are shaded grey.}}      \end{figure*} 
\begin{figure*}[t!]
\centering
\includegraphics{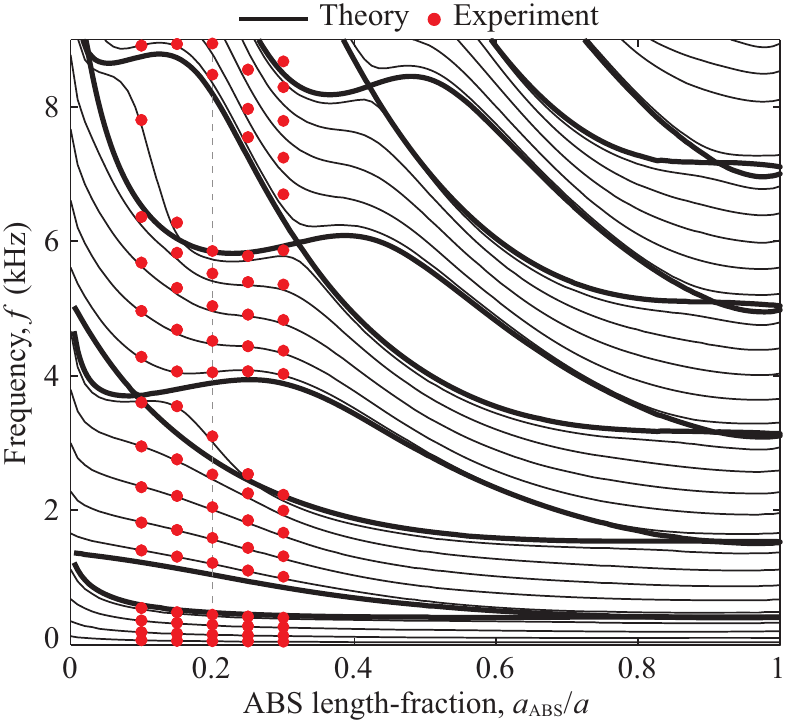}  
\caption{\label{fig:fig12}{Experimental validation: Resonance frequency (thin solid lines, theory; dots, experiment) versus ABS length-fraction for the 5-unit-cell SM-PnC beam structure with $\bar{a} = 203~\text{mm}$. The experimental data points correspond to an ABS length-fraction of 0.1, 0.15, 0.2, 0.25 and 0.3, respectively. The thick solid lines represent the band-gap boundaries for the corresponding infinite periodic materials. }}      \end{figure*} 

Given the localization nature of truncation resonances, the measured amplitude at the far end of the SM-PnC structure is expected to be less than at the edge where the mode is localized and where the excitation is applied. In Fig.~\ref{fig:fig11}, we show using both theory and experiment an FRF comparison between 5- and 6-unit-cell structures in (a) and 5- and 15-unit-cell structures in (b). A truncation resonance peak clearly exists inside the second band gap. We also observe a stronger attenuation from edge-to-edge as the number of unit cells (and total structure length) increases. As for the effect of the number of unit cells on the frequency of the truncation resonance, we note that there is a negligible shift from 5 to 15 unit cells. These results are to be compared with the eigenfrequency versus phason plot shown in Fig.~\ref{fig:fig8}(b) for free-free BCs. It is shown in that figure that beyond 5 unit cells, the change in the frequency of the truncation resonances become negligible. In contrast, the frequencies of the conventional resonances demonstrate substantial shifts, as shown in both Fig.~\ref{fig:fig8}(b) and Fig.~\ref{fig:fig11}. We also observe in Fig.~\ref{fig:fig11}(b) that while the amplitude of the truncation resonance peak drops significantly as the number of unit cells is increased from 5 to 15, the amplitudes of all the conventional resonances do not experience any noticeable drops. 

\subsection{Effect of unit-cell material volume fraction by experiment}
\label{CellVolF}
In addition to property modulation wavenumber and phasons, an alternative approach for controlling the frequency locations of truncation resonances is alternation of the unit-cell design, e.g., by changing its material composition and/or spatial distribution or its geometry. This can result in achieving a total exit of a truncation resoance from a band-gap frequency range, as illustrated in Fig.~\ref{fig:fig12} for a 5-unit-cell SM-PnC structure, which shows that when $a_{\text{ABS}}/a$ is set to 0.25 or higher, no in-gap resonances appear in any of the three gaps covered by both computation and experiment. In this figure, we consider the full range $a_{\text{ABS}}/a$, which at one extreme ($a_{\text{ABS}}/a = 0$) represents a homogenous Al beam, and at the other extreme ($a_{\text{ABS}}/a = 1$) represents a beam composed of only ABS polymer. This figure also allows us to examine the sensitivity of the truncation resonances' frequencies to smooth variations in the material volume fraction. It can be seen that the truncation resonances are noticeably more sensitive to varying the unit-cell layer dimensions than the conventional resonances. Once they exit the band gaps however, these unique resonances become less sensitive to varying $a_{\text{ABS}}/a$, and their sensitivity becomes similar to that of the conventional resonances.

\section{Further reflection on the material vs. structure theme }
\label{Sec:MatvsStr}
The distinction and interconnection between a material and a structure may be examined and classified at various levels. A basic distinction is that of intrinsic versus extrinsic properties or characteristics, e.g., the Young’s modulus and density being intrinsic material properties in contrast to the stiffness and total mass as extrinsic structural characteristics. The distinction may also be made based on physical response. In this context, an elementary classification may be based on the behavior of static deformation, such as the length scale of deformation or spatial span of tangible force interactions. For example, consider a lattice configuration of beams forming a truss that lies at the core of a larger structural frame. If the length scale of deformation at, say, the center of the core is much larger than the individual beam elements and negligible force interaction occurs with the boundaries formed by the frame, then this deformation may be viewed as a form of material behavior. On the other hand, if the length scale of the deformation is on the order of the beam elements, and non-negligible interaction occurs with the boundaries, then the ``periodic network of beams behaves as a structure, such as a frame in a building or a truss in a bridge~\cite{phani2017dynamics}." 

In this work, we have addressed the material-versus-structure correlation problem at a more fundamental level; that is, by examining the characteristics pertaining to finite size in comparison to the properties associated with idealized infinite size, and doing so from a topological elastodynamics  perspective. Here, the dispersion curves represent material properties and the natural frequencies represent structural characteristics. In this context, finite size along the direction where the physical phenomenon of interest takes effect (in this case, wave propagation) is what distinguishes the material versus structure character. Finite dimensions in other lateral dimensions (such as the thickness of a beam, for exmaple) may play a significant role in altering the material properties or structural characteristics, but not in altering the classification of material versus structure. As a periodic material is truncated, and rendered a structure, both bulk and truncation resonances emerge; the latter being intimately connected to the nature of the truncation. This investigation focuses specifically on this aspect. 

\section{Conclusions}
\label{Sec:Conc}
In this paper, we have investigated using theory and experiments the fundamental question of the relation and interplay between material and structure. We provided a formal connection between topological physics and truncation resonances in finite periodic structures. Periodic structures can be understood and topologically characterized using property modulation parameters such as the modulation wavelength $\theta$ and phason $\phi$. These parameters expand the physical space and allow for a rigorous study of the nature of truncation resonances. 

The Chern number is a material property obtained from unit-cell analysis, considering a large number of unit cells with periodic boundary conditions applied. It allows us to predict the behavior of a periodic medium through the bulk-boundary correspondence principle, which in fact is itself a manifestation of the interconnection between the notion of a material and a structure, originated in the quantum realm$-$which we bring here to elastic media. In the QHE theory, for example, the Chern number is a material invariant that predicts the existence of edge currents propagating along the edges of truncated finite samples. Similarly, for our elastic structures, the gap labels predict the number of truncation resonances that span a band gap as $\phi$ is varied for a finite structure with any prescribed BCs.

We have shown that the existence of in-gap truncation resonances cannot be guaranteed for any $\phi$ and that the topological character is understood only when sweeping through $\phi$. This brings a more comprehensive perspective rather than analysing particular truncation cases, and provides a methodology for designing for truncation resonances or their absence. The boundary phasons, which is a concept we introduce in this work, provide an additional tool to control truncation resonances, albeit at different boundaries independently. We have also investigated the effect of the number of unit cells in a finite structure, elucidating that the left- and right-boundary phasons become independent only when a sufficient number of unit cells is present. We similarly demonstrated that the frequency location of truncation resonances converge only when the structure is comprised of a sufficiently large number of unit cells, at least five cells in most cases. Mode couplings$—$whose locations are influenced by the boundary conditions among other factors$—$impact the rate of convergence of the truncation resonances. The impact of the unit-cell constituent material composition was also studied, showing that a truncation resonance may be forced to exit a band gap with an appropriate choice of material volume fraction.

We have also examined another important type of localized mode in finite structures, the defect mode. We have shown it to be non-topological, since it remains flat with change of $\phi$ inside the band gap unless it couples with a truncation resonance. In a perfect ``undefected" periodic structure, there can only be one mode localized at each boundary for any given phason value. By coupling with a defect, it is possible to have two modes localized at the same boundary for a given structure, living inside a band gap, with different frequencies. 

This study, we expect, will inspire future work on multiple fronts. For example, similar principles may be extended to 2D and 3D periodic structures and their truncation resonances, which may manifest as localized modes at points, edges, and surfaces, having connections to topological physics and possibly to higher-order Chern numbers and higher-order topological modes (such as corner modes). Another domain of potential applicability is coiled phononic crystals for space saving~\cite{Willey_2022}. A further angle to be explored in the question of material versus structure is the static regime, where similar connections may be established for topological floppy modes~\cite{Kane_2014}. Other areas to be investigated are the interplay with nonlinearities~\cite{zhou2022topological}, the applicability to damage mechanics such as the effect of number of unit cells on the fracture toughness~\cite{Shaikeea_2022}, and the role of size effects in nanoscience where small finite dimensions have profound impact on thermal transport~\cite{Maruyama_2002} and other physical properties. Implications to quasiperiodic media ~\cite{Liu_2004,pal2019topological,gupta2020dynamics,xia2020topological,rosa2021exploring} or nonperiodic media described statistically by representative volume elements may also be explored. Finally, the framework presented for connecting between topology and truncation may potentially be applied to finite systems in other branches of physics, such as photonics~\cite{yeh1978optical} and quantum mechanics~\cite{ren2010surface}.

\section*{Acknowledgement}
The authors acknowledge the students Andrew S. Tomchek and Edgar A. Flores for their assistance in conducting the experiments. 


\bibliographystyle{elsarticle-num-names}
\bibliography{litrev_JMPS}



\end{document}